\documentclass[conference]{IEEEtran}
\IEEEoverridecommandlockouts
\usepackage{amsmath,amsfonts}
\usepackage{algorithmic}
\usepackage{algorithm}
\usepackage{array}
\usepackage{textcomp}
\usepackage{stfloats}
\usepackage{url}
\usepackage{verbatim}
\usepackage{graphicx}
\usepackage{cite}
\usepackage{multirow}
\usepackage{booktabs}
\usepackage{subfigure}
\usepackage[dvipsnames]{xcolor}
\usepackage{bbm}
\usepackage{enumitem}
\usepackage{marvosym}
\def\BibTeX{{\rm B\kern-.05em{\sc i\kern-.025em b}\kern-.08em
    T\kern-.1667em\lower.7ex\hbox{E}\kern-.125emX}}
\begin{document}

\title{Efficient Model-Agnostic Continual Learning for Next POI Recommendation}

\author{
Chenhao Wang\textsuperscript{1}, Shanshan Feng\textsuperscript{2}, Lisi Chen\textsuperscript{1}, Fan Li\textsuperscript{3}, Shuo Shang\textsuperscript{1}\textsuperscript{\Letter}\thanks{\Letter~Corresponding author: Shuo Shang.} \\
\textsuperscript{1}University of Electronic Science and Technology of China, Chengdu, China \\ \textsuperscript{2}Wuhan University, Wuhan, China \quad
\textsuperscript{3}The Hong Kong Polytechnic University, Hong Kong, China \\
chenhao.wang@std.uestc.edu.cn, victor\_fengss@whu.edu.cn, lchen012@e.ntu.edu.sg \\ fan-5.li@polyu.edu.hk, jedi.shang@gmail.com\\
}

\maketitle
\begin{abstract}
Next point-of-interest (POI) recommendation improves personalized location-based services by predicting users’ next destinations based on their historical check-ins. However, most existing methods rely on static datasets and fixed models, limiting their ability to adapt to changes in user behavior over time. To address this limitation, we explore a novel task termed continual next POI recommendation, where models dynamically adapt to evolving user interests through continual updates. This task is particularly challenging, as it requires capturing shifting user behaviors while retaining previously learned knowledge. Moreover, it is essential to ensure efficiency in update time and memory usage for real-world deployment. To this end, we propose GIRAM (Generative Key-based Interest Retrieval and Adaptive Modeling), an efficient, model-agnostic framework that integrates context-aware sustained interests with recent interests. GIRAM comprises four components: (1) an interest memory to preserve historical preferences; (2) a context-aware key encoding module for unified interest key representation; (3) a generative key-based retrieval module to identify diverse and relevant sustained interests; and (4) an adaptive interest update and fusion module to update the interest memory and balance sustained and recent interests. In particular, GIRAM can be seamlessly integrated with existing next POI recommendation models. Experiments on three real-world datasets demonstrate that GIRAM consistently outperforms state-of-the-art methods while maintaining high efficiency in both update time and memory consumption.
\end{abstract}


\section{INTRODUCTION}
With the proliferation of location-based services, next point-of-interest (POI) recommendation has become a key technology for improving user experience across various applications~\cite{DBLP:journals/tkde/ZhangYYWHYY25}, including navigation systems~\cite{DBLP:conf/webi/RutaSIFS15}, tourism platforms~\cite{DBLP:journals/tkde/YangDWHXS19}, and food delivery services~\cite{DBLP:conf/gis/SongLCCHS21}. Accurate recommendations rely on effectively modeling user preferences with historical check-in data. Deep learning models, such as RNNs~\cite{DBLP:conf/aaai/LiuWWT16,DBLP:conf/ijcai/Kong018,DBLP:conf/www/FengLZSMGJ18,DBLP:conf/aaai/SunQCLNY20,DBLP:conf/ijcai/YangFRC20}, Transformers~\cite{DBLP:conf/sigir/YangL022,DBLP:conf/kdd/FengMCSO24}, and GNNs~\cite{DBLP:conf/kdd/RaoCLSYH22,DBLP:conf/sigir/YanSJHWLC23} have been widely adopted to capture sequential dependencies and spatio-temporal patterns in user trajectories.

Despite their effectiveness, most existing methods follow a static training paradigm: models are trained once on historical data, and then remain fixed during deployment. Consequently, these models fail to adapt to users' evolving interests and cannot incorporate new check-in data as it becomes available. This static paradigm conflicts with the dynamic nature of user interests, which can change due to multiple factors such as time, location, and POI categories. For example, a user may prefer indoor venues (e.g., coffee shops and museums) in winter and outdoor places (e.g., parks and beaches) in summer. These shifts highlight the need for recommendation methods that can continually adapt to user behavior.

\begin{figure}[!t]
\centering
\includegraphics[width=0.9\linewidth, trim=0 13 5 10, clip]{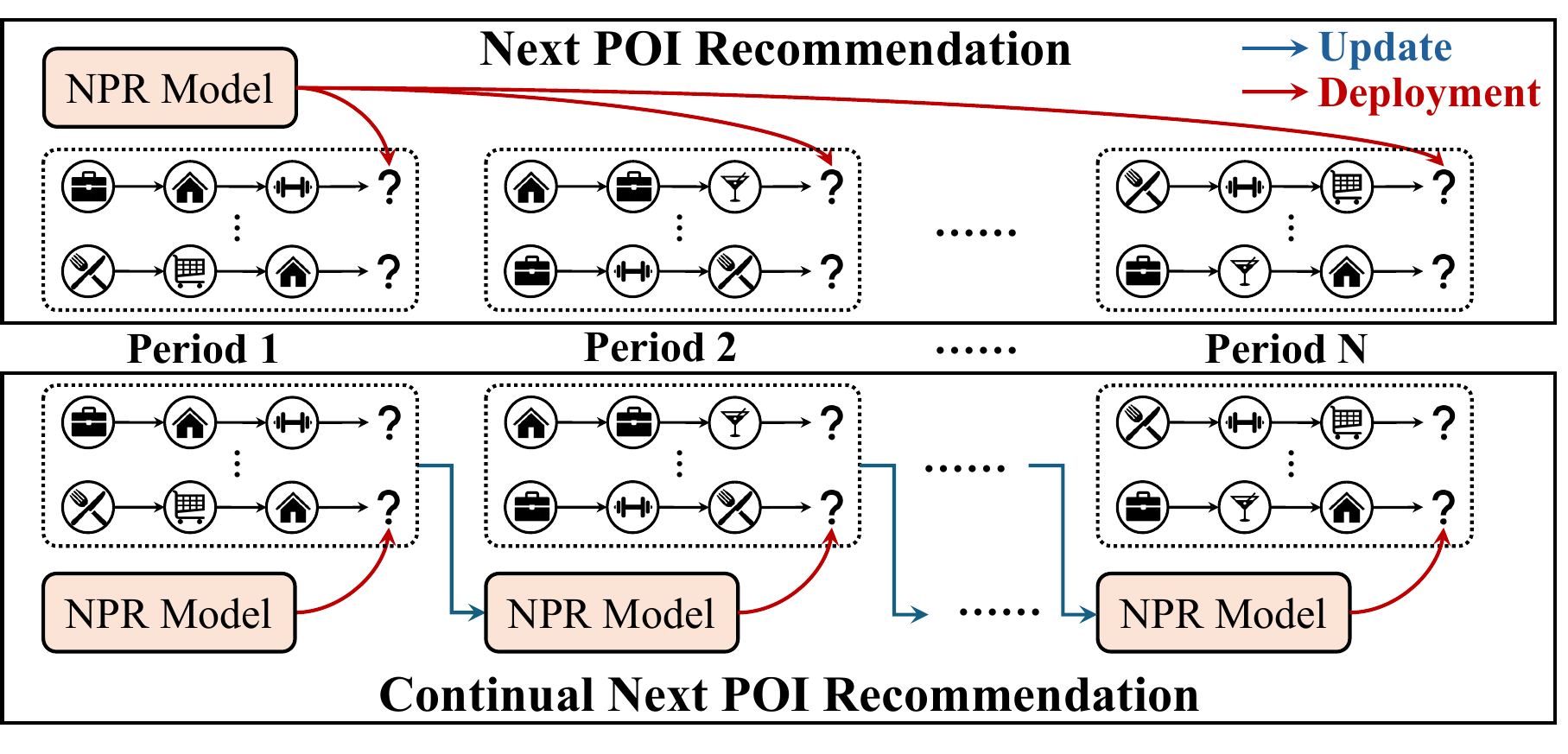}
\vspace{-9pt}
\caption{Illustration of traditional vs. continual next POI recommendation.}
\vspace{-18pt}
\label{fig:intro}
\end{figure}

\begin{figure}[!t]
\centering
\vspace{-4pt}
\includegraphics[width=1\linewidth]{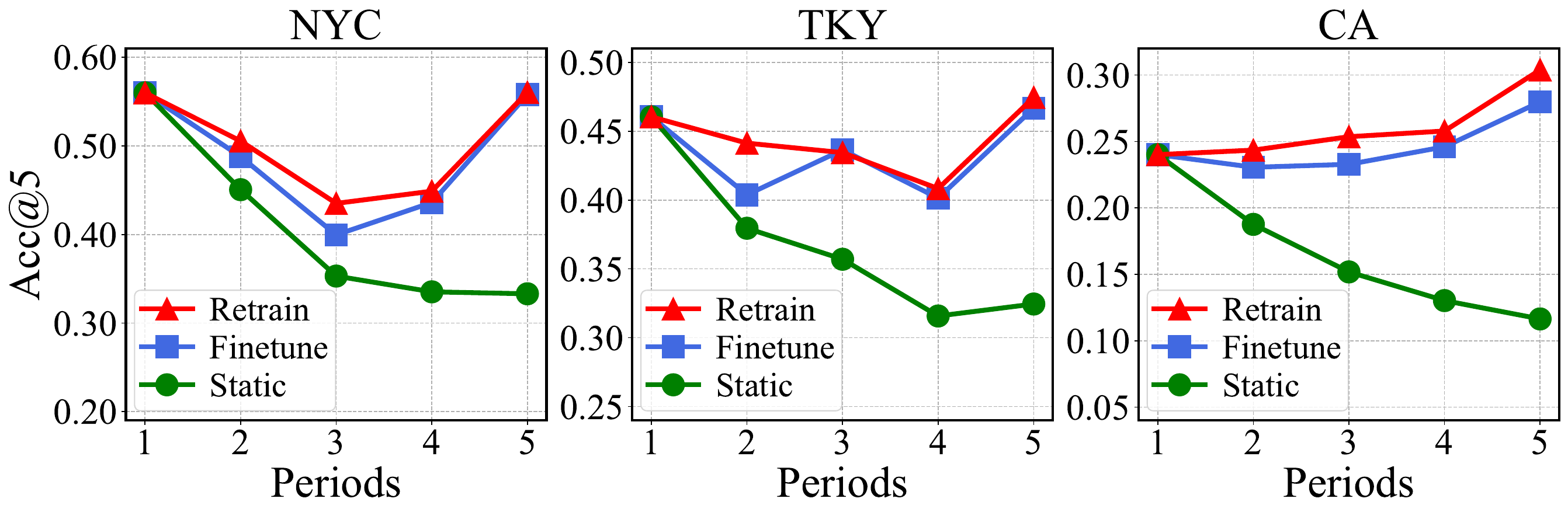}
\vspace{-20pt}
\caption{Performance comparison of Static, Finetune, and Retrain methods on NYC, TKY, and CA datasets.}
\vspace{-18pt}
\label{fig:data_observation}
\end{figure}

Therefore, in this study, we investigate the novel problem of continual next POI recommendation. As shown in Figure~\ref{fig:intro}, traditional next POI recommendation neglects temporal interest shifts and uses a statically trained model with fixed parameters during deployment. In contrast, continual next POI recommendation updates the model periodically to capture evolving user interests, as illustrated by the blue arrows. Furthermore, Figure~\ref{fig:data_observation} shows that the performance of static models degrades over time, underscoring the necessity for continual model updates. One intuitive solution is to periodically retrain the model using all observed data, but this is computationally expensive. Another straightforward approach is to finetune the model using only recently observed data. However, this can result in catastrophic forgetting, where previously learned knowledge is overwritten, resulting in lower performance than retraining. To overcome this, the continual next POI recommendation aims to dynamically adapt models to evolving user interests while maintaining historical knowledge through efficient updates without full retraining.

However, designing an effective and efficient framework for continual next POI recommendation presents several key challenges: \textbf{(1) Preserving historical preferences.} The framework must retain prior knowledge while adapting to new user behaviors to avoid catastrophic forgetting. This is challenging due to the context-sensitive nature of human mobility, where user behavior is shaped by spatio-temporal and categorical factors. \textbf{(2) Retrieving relevant interests.} As user interests evolve, not all historical preferences remain relevant. Therefore, the system must selectively retrieve useful information from historical data. Effective retrieval ensures that sustained interests, those persist across time and context, are accurately identified. This requires prioritizing relevant information across factors such as location, time, and activity, while minimizing redundancy and information loss. \textbf{(3) Balancing historical and recent information.} High-quality continual recommendations require integrating both sustained and recent interests. Sustained interests provide a foundation for modeling long-term behavior, while recent interests capture immediate needs. Balancing these interests is difficult, as the relative importance of historical and current information varies across users and contexts. \textbf{(4) Ensuring model-agnostic adaptability.} A continual recommendation framework should integrate with diverse NPR models, such as RNNs~\cite{DBLP:conf/ijcai/YangFRC20}, Transformers~\cite{DBLP:conf/sigir/YangL022}, and diffusion models~\cite{DBLP:journals/tois/QinWJLZ24}, without architecture modifications. However, achieving such adaptability while maintaining performance is challenging, due to the heterogeneous design characteristics of different models.

To address these challenges, GIRAM (\underline{G}enerative Key-based \underline{I}nterest \underline{R}etrieval and \underline{A}daptive \underline{M}odeling) is proposed as a novel model-agnostic framework for continual next POI recommendation, which integrates context-relevant historical and current information. The framework comprises four key components. For the first challenge, we construct an \emph{interest memory} to preserve historical user preferences. This memory stores contextual representations as keys and the corresponding sparse prediction outputs from the finetuned NPR model as values. Unlike raw data storage, this method retains abstract behavioral patterns while remaining efficient and scalable. For the second challenge, we design a \emph{generative key-based retrieval} module that generates multiple candidate keys to query the interest memory and capture the diversity of user interests across contexts. Retrieving with multiple keys enables comprehensive access to sustained interests as a single key may not capture complex contextual information. For the third challenge, we design an \emph{adaptive interest update and fusion} module that uses a consistency score to dynamically balance historical and current information. This mechanism improves interest memory updates and recommendations by adapting to shifts in user behavior despite limited memory size. For the fourth challenge, a model-agnostic \emph{context-aware key encoding} module is designed to encode contextual information into unified key vectors. This lightweight design enables seamless and low-overhead integration with various NPR models.

The continual learning process of GIRAM consists of two stages: the \emph{update} stage and the \emph{deployment} stage. In the update stage, the context-aware key encoding module generates interest keys by encoding spatio-temporal and categorical features into unified key vectors. These keys are paired with outputs from the finetuned NPR model to form structured user preferences stored in the interest memory. A consistency score is introduced to adaptively balance historical and new preferences during updates. In the deployment stage, the generative key-based retrieval module uses a conditional VAE model to generate multiple similar but independent keys to retrieve sustained interests from the interest memory. The sustained and recent interests are then fused with the consistency score adaptively balancing their contributions. This approach allows efficient continual adaptation without full model retraining while effectively using previously learned knowledge.

Our main contributions can be summarized as follows:
\begin{itemize}[leftmargin=*]
\item A novel research problem, continual next POI recommendation, is introduced. This problem considers dynamic shifts in user interests and continual model updates, making it more challenging than traditional next POI recommendation.
\item We propose GIRAM, an efficient model-agnostic framework that integrates interest memory, context-aware key encoding, generative key-based retrieval, and adaptive interest update and fusion. To the best of our knowledge, this is the first study to systematically address the problem of continual next POI recommendation.
\item GIRAM can be seamlessly integrated with various existing next POI recommendation models, efficiently enhancing their performance in continual learning settings without requiring architectural modifications or full retraining.
\item Extensive experiments on three real-world datasets show that GIRAM outperforms state-of-the-art baselines while maintaining high efficiency in both update time and memory consumption. Our data and source code are available at: \url{https://github.com/chwang0721/GIRAM}.
\end{itemize}

\section{RELATED WORK}
\label{sec:related_work}
\vspace{-2pt}
\subsection{Next POI Recommendation}
\vspace{-1pt}
Next POI recommendation is crucial in location-based services, and various methods have been proposed. Early approaches~\cite{DBLP:conf/www/RendleFS10,DBLP:conf/ijcai/FengLZCCY15,han2019auc,han2020contextualized,han2021point} use matrix factorization, while sequential models such as ST-RNN~\cite{DBLP:conf/aaai/LiuWWT16} and HST-LSTM~\cite{DBLP:conf/ijcai/Kong018} integrate spatial and temporal contexts through RNNs. Attention mechanisms enhance trajectory modeling in models like DeepMove~\cite{DBLP:conf/www/FengLZSMGJ18}, LSTPM~\cite{DBLP:conf/aaai/SunQCLNY20}, Flashback~\cite{DBLP:conf/ijcai/YangFRC20}, STAN~\cite{DBLP:conf/www/LuoLL21}, and CFPRec~\cite{DBLP:conf/ijcai/ZhangSW0OQ22}. GNNs further advance recommendations by modeling complex relationships in check-in data. STP-UDGAT~\cite{DBLP:conf/cikm/LimHNWGWV20} applies graph attention for global contexts, while HMT-GRN~\cite{DBLP:conf/sigir/LimHNGWT22} handles data sparsity with multi-task graph recurrent networks. DRGN~\cite{DBLP:conf/sigir/WangZLW22} combines graph-based representations with trajectory flow maps, and Graph-Flashback~\cite{DBLP:conf/kdd/RaoCLSYH22} integrates spatial-temporal knowledge graphs with sequential models. STHGCN~\cite{DBLP:conf/sigir/YanSJHWLC23} uses hyper-graph convolution for diverse user behaviors, while SNPM~\cite{DBLP:conf/aaai/Yin00CS023} and AGCL~\cite{DBLP:journals/tkde/RaoJSCHYK25} employ dynamic graphs for POI relations. Recent innovations include ROTAN~\cite{DBLP:conf/kdd/FengMCSO24} with rotation-based attention, DiffPOI~\cite{DBLP:journals/tois/QinWJLZ24} with diffusion models, LLM4POI~\cite{DBLP:conf/sigir/LiR0A0S24} and GNPR-SID~\cite{wang2025generative} leveraging large language models for enhanced recommendations.

However, most existing methods rely on static data, ignoring the evolution of user interests. As user check-ins are continuously generated, periodically retraining on all data is computationally expensive, while finetuning on recent data leads to catastrophic forgetting. Our goal is to develop models that adapt to new data while effectively retaining prior knowledge.

\subsection{Continual Learning for Recommendation}
\vspace{-1pt}
Continual learning enables models to adapt to shifting data distributions~\cite{DBLP:journals/pami/WangZSZ24}, making it essential for recommendation systems with evolving user preferences~\cite{DBLP:conf/sigir/WangYHWDN18}. Existing continual learning methods in recommendation can be broadly classified into replay-based methods and model-based methods.

Replay-based methods update models by replaying historical interactions. Ranking-based methods~\cite{DBLP:conf/recsys/Diaz-AvilesDSN12,DBLP:journals/pvldb/ChenYYC13,DBLP:conf/sigir/WangYHWDN18} select fixed-size positive and negative samples for updates. Session-based methods prioritize sessions with lower model performance~\cite{DBLP:conf/kdd/GuoYWCZH19,DBLP:conf/sigir/QiuYHC20} or sample based on item frequency~\cite{DBLP:conf/recsys/MiLF20}. MAN~\cite{DBLP:conf/ijcai/MiF20} retrieves relevant interactions from nonparametric memory, and ReLoop2~\cite{DBLP:conf/kdd/ZhuCHDT023} refines predictions using similar samples and error memory. Model-based methods preserve historical knowledge from trained models. Some adopt knowledge distillation~\cite{DBLP:journals/corr/abs-2009-02147,DBLP:conf/cikm/XuZGGTC20,DBLP:conf/icde/WangS23,DBLP:conf/kdd/LeeKKY24}, while others apply meta-learning~\cite{DBLP:conf/sigir/ZhangFW00LZ20,DBLP:journals/geoinformatica/LvSTCSQCZ23,DBLP:conf/recsys/PengPZZ21}. FIRE~\cite{DBLP:conf/www/XiaLGLLG22} updates the interaction matrix via graph signal processing, and CPMR~\cite{DBLP:conf/cikm/BianXFK23} captures historical and contextual dynamics via a pseudo-multi-task framework.

Despite these advances, most continual recommendation methods focus on e-commerce and media and are not directly applicable to continual next POI recommendation due to their limited ability to model spatio-temporal dependencies and integrate with existing NPR models. In contrast, continual next POI recommendation, which is characterized by complex spatio-temporal patterns, remains largely underexplored.

\subsection{Spatio-temporal Continual Learning}
\vspace{-1pt}
Recent works have explored continual learning for spatio-temporal prediction tasks, such as traffic flow and vehicle speed prediction. URCL~\cite{DBLP:conf/icde/00010GY0HXJ24} mitigates catastrophic forgetting by integrating a replay buffer with historical samples and designing a spatio-temporal mix-up mechanism. CMuST~\cite{DBLP:conf/nips/YiZHCYW024} dissects spatio-temporal interactions through its Rolling Adaptation training scheme, balancing task-specific features and shared patterns across tasks. DOST~\cite{DBLP:journals/corr/abs-2411-15893} employs an adaptive spatio-temporal network with a Variable-Independent Adapter for location-specific adaptation, addressing the challenge of gradual data distribution shifts. STONE~\cite{DBLP:conf/kdd/WangMWW0ZW24} enhances traffic prediction generalization using Fréchet embeddings and semantic graph perturbations to capture stable node dependencies. UFCL~\cite{DBLP:journals/tkde/MiaoZGYZJ25}, an extension of URCL, addresses decentralized, privacy-sensitive data through federated learning.

However, these methods are unsuitable for continual next POI recommendation due to differences between continuous traffic data and discrete check-in data. URCL and UFCL rely on linear combinations of observations, CMuST averages observations over equivalent times of day, STONE generalizes node dependencies via graph perturbations, and DOST addresses unique distribution shifts at each location. These approaches are designed for continuous traffic streams and cannot handle discrete POI data or user-specific information, which are crucial for personalized next POI recommendation.
\section{PRELIMINARIES AND ANALYSIS}

\subsection{Problem Definition}
\vspace{-1pt}
In this section, we introduce the necessary definitions and state the continual next POI recommendation problem.

Let $\mathcal{U} = \{u_1, u_2, \dots, u_{|\mathcal{U}|}\}$ denote the set of users, and $\mathcal{P} = \{p_1, p_2, \dots, p_{|\mathcal{P}|}\}$ denote the set of POIs. Each POI is characterized by two attributes: its geographic coordinate and its category. A check-in record is represented as a tuple $r = (u, p, t)$, which denotes that user $u$ visits POI $p$ at timestamp $t$. By partitioning each user's check-in records based on a given time interval, we construct check-in trajectories. A trajectory of length $n$ is represented as $T = \langle r_1, r_2, \dots, r_n \rangle$, where each element $r_i$ corresponds to a particular check-in record.

\textbf{Next POI Recommendation (NPR)}. Given a set of historical check-in trajectories $\mathcal{T} = \{T_1, T_2, \dots, T_{|\mathcal{T}|}\}$, and a current trajectory $T_{\text{curr}} = \langle r_1, r_2, \dots, r_m \rangle$ of a user $u$, the goal of next POI recommendation is to predict the next POI $p_{m+1}$ that the user is most likely to visit, based on the historical trajectories in $\mathcal{T}$ and the check-in records in the current trajectory $T_{\text{curr}}$.

\textbf{Continual Next POI Recommendation (CNPR)}. Given a data stream $[\mathcal{T}_1, \dots, \mathcal{T}_k, \mathcal{T}_{k+1}]$, where $\mathcal{T}_i$ denotes the block of check-in trajectories collected during period $i$. The objective of the continual next POI recommendation in period $k$ is to update the NRP model and perform recommendations for the trajectories in the upcoming data block $\mathcal{T}_{k+1}$. The key feature of CNPR is to continually update the recommendation system as new data blocks become available. During period $k$, the system processes the current block $\mathcal{T}_k$, and leverages the accumulated knowledge obtained from all previous data blocks $[\mathcal{T}_1, \mathcal{T}_2, \dots, \mathcal{T}_{k-1}]$. This ensures that the predictions for the next data block $\mathcal{T}_{k+1}$ are both informed by historical data and adapted to the latest user behavior patterns.

\vspace{-1pt}
\subsection{Data Observations and Analyses}
\vspace{-1pt}
To illustrate our motivation, we analyze the performance of static, finetuned, and retrained models using Flashback~\cite{DBLP:conf/ijcai/YangFRC20} as the backbone (experimental setup detailed in Section~\ref{sec:5.1}). The results shown in Figure~\ref{fig:data_observation} present three key observations: (1) The performance of static models degrades over time as user preferences evolve, indicating their inability to adapt to dynamic check-in data. (2) The performance gap between static models and finetuned or retrained models widens over time, underscoring the need for continual model updates (e.g., the gap in the 5th period is much larger than the 2nd). (3) Retrained models outperform finetuned ones, as finetuning solely on new data leads to catastrophic forgetting. This highlights the necessity of integrating historical and newly collected data. However, retraining on entire datasets periodically is computationally expensive and impractical in real-world applications. These observations underscore the need for a framework that preserves historical user interests while balancing performance and efficiency, without requiring frequent retraining.
\section{METHODOLOGY}
\begin{figure*}[!t]
\centering
\vspace{-4pt}
\includegraphics[width=0.98\textwidth,trim=0 12 12 10,clip]{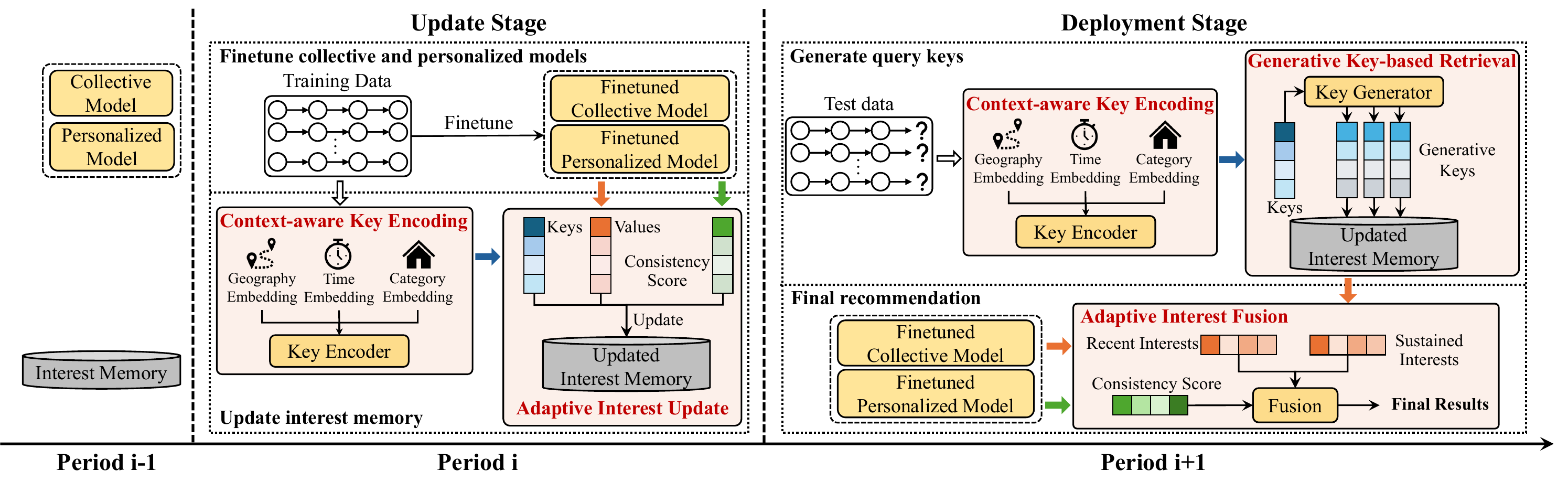}
\vspace{-10pt}
\caption{An overview of GIRAM. It can be divided into two stages: the update stage and the deployment stage.}
\vspace{-16pt}
\label{fig:framework}
\end{figure*}

\vspace{-1pt}
Figure~\ref{fig:framework} shows the overall framework of our proposed GIRAM. The process of continual next POI recommendation using GIRAM is iteratively decomposed into two stages: the \emph{update} stage and the \emph{deployment} stage. The update stage preserves and updates historical user preferences. The deployment stage retrieves sustained interests and integrates them adaptively with recent interests to generate recommendations.

GIRAM comprises four core components: (1) interest memory; (2) context-aware key encoding; (3) generative key-based retrieval; and (4) adaptive interest update and fusion. The interest memory preserves historical user preferences in a key-value format, where keys are generated by the context-aware encoding module, and values are derived from the finetuned NPR model outputs. In the update stage, the key encoding module encodes the contextual information from the training data into unified keys, which are then used by the interest update module to refresh the memory. During deployment, the keys extracted from the test data serve as queries to retrieve sustained interests via the generative key-based retrieval module. These are then fused with recent interests predicted by the finetuned NPR model using the interest fusion module, producing a comprehensive representation of user interests. The following sections detail each component of GIRAM.

\vspace{-1pt}
\subsection{Interest Memory}
\vspace{-1pt}
In the CNPR problem, retraining from scratch can be computationally inefficient and may introduce redundant or irrelevant information, potentially degrading performance. To address this, our goal is to update the model using only newly collected check-ins while retaining previously acquired knowledge. This section presents an interest memory designed to preserve historical user preferences effectively and efficiently.

The proposed interest memory adopts a user-specific structure, denoted by $M = \{M_u\}_{u=1}^{|\mathcal{U}|}$, each user's memory $M_u$ is defined as a set of key-value-timestamp triples:
\begin{equation}
\vspace{-1pt}
M_u = \{(\mathbf{k}_i, \mathbf{v}_i, t_i)\ |\,\ 1 \leq i \leq N_u \},
\vspace{-1pt}
\end{equation}
where $N_u$ is the memory capacity for user $u$, $\mathbf{k}_i \in \mathbb{R}^{d_k}$ is the key representation computed by the Context-aware Key Encoding module (see Section~\ref{sec:key_encoding}), the value vector $\mathbf{v}_i \in \mathbb{R}^{d_v}$ is the predicted POI distribution, and $t_i$ is the timestamp. 

Unlike prior approaches that use discrete labels as memory values~\cite{DBLP:conf/ijcai/MiF20,DBLP:conf/kdd/ZhuCHDT023}, we adopt a distributional representation, where each value vector $\mathbf{v}_i$ stores the top-$K$ prediction probabilities of POIs, obtained from the output of a finetuned next POI recommendation model $\mathcal{V}$:
\begin{equation}
\vspace{-1pt}
\mathbf{v}_i = \mathrm{Top}_K \bigl(\mathrm{Softmax}(\mathcal{V}_{\theta}(T_i)) \bigr),
\vspace{-1pt}
\end{equation}
where $\mathcal{V}_{\theta}(T_i)$ denotes the raw output of the NPR model with parameters $\theta$ given input $T_i$, and $\mathrm{Top}_K(\cdot)$ retains $K$ largest values in the vector and sets the remaining dimensions to zero, which is then stored using a sparse matrix. This design has two main advantages. First, it captures complex contextual dependencies, providing richer, more accurate representations of dynamic user interests than discrete labels. Second, it increases robustness to label noise and overfitting by modeling generalizable patterns while maintaining memory efficiency.

\vspace{-1pt}
\subsection{Context-aware Key Encoding}
\vspace{-1pt}
\label{sec:key_encoding}
Existing continual recommendation methods often use the hidden states of recommendation models as query representations for memory retrieval or experience replay~\cite{DBLP:conf/recsys/MiLF20,DBLP:conf/kdd/ZhuCHDT023}. However, this strategy presents two main drawbacks. First, hidden states tend to be tightly coupled with model outputs, which are often linear transformations of these states. This coupling introduces redundancy between memory keys and values, reducing the effectiveness of retrieval and limiting the model’s ability to capture diverse user interests. Second, the quality of hidden states depends heavily on model performance, so suboptimal training can may produce poor context representations, compromising retrieval. 
% To address these issues, we propose a model-agnostic context-aware key encoding module that produces unified key representations independently of the recommendation outputs.
To address these issues, a model-agnostic context-aware key encoding module is developed to produce unified key representations independently of the recommendation outputs.

\subsubsection{Geography Embedding}
To capture spatial features, we combine continuous GPS coordinates with discrete region identifiers. The GPS coordinates are first normalized via min-max scaling and then projected into a latent spatial space through a nonlinear transformation:
\begin{equation}
\vspace{-1pt}
\mathbf{l}_\text{norm} = \mathcal{G}_{\text{geo}}(l_{\text{lat}}, l_{\text{lon}}), \quad  \mathbf{\Theta}_{\text{coord}} = \mathbf{W}_{\text{coord}} \cdot \mathbf{l}_\text{norm} + \mathbf{b}_{\text{coord}},
\end{equation}
where $\mathcal{G}_{\text{geo}}(\cdot)$ denotes the min-max normalization function applied to the latitude and longitude, and $\mathbf{W}_{\text{coord}}$, $\mathbf{b}_{\text{coord}}$ are trainable parameters that map the normalized coordinates into a coordinate embedding space.

To model higher-level spatial semantics, we further discretize the coordinate space into a uniform grid following~\cite{DBLP:conf/icde/LiZCJW18}, assigning each location a region identifier $g$. The corresponding region embedding is computed as:
\begin{equation}
\vspace{-1pt}
\mathbf{\Theta}_{\text{region}} = \mathcal{E}_{\text{region}}(g),
\vspace{-1pt}
\end{equation}
where $\mathcal{E}_{\text{region}}(\cdot)$ is an embedding function that maps the discrete region ID to a dense vector via a learnable embedding layer. Finally, the complete geography embedding is formed by concatenating the coordinate and region embeddings:
\begin{equation}
\vspace{-1pt}
\mathbf{\Theta} = \mathbf{\Theta}_{\text{coord}} \parallel \mathbf{\Theta}_{\text{region}},
\vspace{-1pt}
\end{equation}
where $(\cdot\parallel\cdot)$ denotes concatenation operation. This approach effectively combines fine-grained spatial details with coarse-grained regional semantics, thereby enhancing the expressiveness of spatial representations.

\subsubsection{Time Embedding}
The time embedding module combines discrete temporal attributes and continuous periodic patterns to model temporal contexts. The discrete temporal embedding is defined as follows:
\begin{equation}
\vspace{-1pt}
\mathbf{\Phi}_{\text{discrete}} = \mathcal{E}_{\text{hour}}(h) \parallel \mathcal{E}_{\text{weekday}}(w),
\vspace{-1pt}
\end{equation}
where $h$ is the hour (from 0 to 23), and $w$ is the weekday (from 0 for Monday to 6 for Sunday). $\mathcal{E}_{\text{hour}}(\cdot)$ and $\mathcal{E}_{\text{weekday}}(\cdot)$ are embedding layers mapping discrete values to latent vectors. In addition to discrete encoding, we incorporate periodic positional encoding to model cyclic temporal patterns. The normalized time of day $\tau \in [0, 1]$ is computed as the current time in seconds divided by the total number of seconds in a day. Based on this, the periodic embedding is formulated as:
\begin{equation}
\vspace{-1pt}
\mathbf{\Phi}_{\text{periodic}} = [\sin(2 \pi \omega \tau) \parallel \cos(2 \pi \omega \tau) \,|\, \omega \in \boldsymbol{\mathcal{F}}],
\vspace{-1pt}
\end{equation}
where $\boldsymbol{\mathcal{F}}$ is a set of frequencies (e.g., ${1, 2, 4}$) used to capture multi-scale periodic patterns. The final time embedding is obtained by concatenating the discrete and periodic components:
\begin{equation}
\vspace{-1pt}
\mathbf{\Phi} = \mathbf{\Phi}_{\text{discrete}} \parallel \mathbf{\Phi}_{\text{periodic}}.
\vspace{-1pt}
\end{equation}

This design effectively combines discrete and periodic temporal features, enabling the model to capture both specific time information and periodic patterns.

\subsubsection{Category Embedding}
The third component is the category embedding, which is crucial for next POI recommendation, as POI categories reflect user activity types. Using only the original category ID often fails to capture semantic relationships among related POIs. To address this, we incorporate both raw and derived category information. The raw category corresponds to the original category label assigned to the POI, while the derived category is inferred from contextual semantics using ChatGPT, following the methodology proposed in LLM4POI~\cite{DBLP:conf/sigir/LiR0A0S24}. For example, in the NYC dataset, categories such as "Food Truck", "Restaurant", and "Burger Joint" are unified under the derived category "Food and Dining", capturing broader semantic groupings. The embeddings for raw and derived categories are computed as:
\begin{equation}
\vspace{-1pt}
\mathbf{\Omega} = \mathcal{E}_{\text{raw}}(c_{\text{raw}}) \parallel \mathcal{E}_{\text{der}}(c_{\text{der}}),
\vspace{-1pt}
\end{equation}
where $c_{\text{raw}}$ and $c_{\text{der}}$ denote the raw and derived category IDs, respectively. $\mathcal{E}_{\text{raw}}(\cdot)$ and $\mathcal{E}_{\text{der}}(\cdot)$ are embedding layers that project these categorical features into latent semantic spaces.

\subsubsection{Key Encoder}
Finally, the key encoder combines the above embeddings into a unified representation and then processes it through a sequence model to capture temporal dependencies. The combined representation is formulated as:
\begin{equation}
\vspace{-1pt}
\mathbf{k} = \mathrm{LSTM}\bigl(\mathrm{Linear}(\mathbf{\Theta} \parallel \mathbf{\Phi} \parallel \mathbf{\Omega})\bigr),
\vspace{-1pt}
\end{equation}
where $\mathbf{\Theta}$ is the geography embedding, $\mathbf{\Phi}$ is the time embedding, and $\mathbf{\Omega}$ is the category embedding. The output $\mathbf{k}$ is the final key representation, which will be used for both interest memory update and generative key-based retrieval.

\subsection{Generative Key-based Retrieval}
With interest memory, a straightforward retrieval method is to find the most similar key and use its value directly. However, a single key cannot capture the diversity of user interests across contexts. 
% To this end, we propose a generative key-based retrieval module that generates multiple relevant yet independent keys. 
To this end, a generative key-based retrieval module is designed to generate multiple relevant yet independent keys.
Specifically, a conditional VAE-based model is employed to produce keys, which are used to retrieve diverse sustained interests from the interest memory. This strategy ensures broader representations, enabling more comprehensive recommendations by balancing relevance and diversity.

\subsubsection{Key Generator}
The key generator adopts a conditional VAE architecture consisting of an encoder $\mathcal{Q}_{\phi}$, a latent sampler, and a decoder $\mathcal{P}_{\psi}$. This design enables the generation of diverse yet contextually relevant query keys via latent space sampling. Unlike deterministic methods (e.g., autoencoders), which cannot model uncertainty, and adversarial methods (e.g., GANs), which often suffer from instability and mode collapse, the VAE offers a robust balance between generative diversity and training stability. Given an input context key $\mathbf{k}$, the encoder estimates a posterior distribution over latent variables:
\begin{equation}
\begin{split}
\mathcal{Q}_{\boldsymbol{\phi}}(\mathbf{z} \mid \mathbf{k}) = \mathcal{N}(\boldsymbol{\mu}, \mathrm{diag}(\boldsymbol{\sigma}^2)), \\
\boldsymbol{\mu} = f_{\mu}(\mathbf{k}),\quad 
\log \boldsymbol{\sigma}^2 = f_{\sigma}(\mathbf{k})
\end{split}
\end{equation}
where $\boldsymbol{\mu}$ and $\boldsymbol{\sigma}^2$ are the mean and variance of the Gaussian posterior, computed by two separate MLPs $f_{\mu}$ and $f_{\sigma}$. A latent vector $\mathbf{z}$ is then sampled via the reparameterization trick:
\begin{equation}
\label{eq:reparam}
\vspace{-1.5pt}
\mathbf{z} = \boldsymbol{\mu} + \boldsymbol{\epsilon} \odot \boldsymbol{\sigma},\quad \boldsymbol{\epsilon} \sim \mathcal{N}(\mathbf{0}, \mathbf{I}),
\vspace{-1.5pt}
\end{equation}

To generate $N_k$ diverse keys, multiple latent vectors $[\mathbf{z}_1, \dots, \mathbf{z}_{N_k}]$ are sampled from Equation~\ref{eq:reparam}. Each is conditioned by the original key $\mathbf{k}$ and passed through a decoder:
\begin{equation}
\mathbf{k}_i = f_{\text{act}}\bigl(\mathcal{P}_{\psi}(\mathbf{z}_i, \mathbf{k})\bigr),
\end{equation}
where $\mathcal{P}_{\psi}$ is a decoder MLP parameterized by $\psi$,
and $f_{\text{act}}(\cdot)$ denotes a Sigmoid activation. This generates a set of keys $\mathbf{K} = [\mathbf{k}_1, \dots, \mathbf{k}_{N_k}]$. To ensure alignment with the original key, a reconstruction loss minimizes the discrepancy between the original and the generated keys:
\begin{equation}
\vspace{-1.5pt}
\mathcal{L}_{\text{recon}} = \left \| \mathbf{k} - \frac{1}{N_k} \sum_{i=1}^{N_k} \mathbf{k}_{i} \right \|^2.
\vspace{-1.5pt}
\end{equation}

As in standard VAEs, a KL divergence term is used to regularize the posterior toward the standard Gaussian prior:
\begin{equation}
\mathcal{L}_{\text{KL}} = D_{\text{KL}}\bigl( \mathcal{Q}_{\phi}(\mathbf{z} \,|\, \mathbf{k}) \,\|\, \mathcal{N}(\mathbf{0}, \mathbf{I}) \bigr).
\end{equation}

To further encourage key diversity, a pairwise diversity loss penalizes redundancy by minimizing the inverse squared distances between key pairs:
\begin{equation}
\vspace{-1.5pt}
\mathcal{L}_{\text{div}} = \sum_{i=1}^{N_k} \sum_{j=i+1}^{N_k} \frac{1}{\left\| \mathbf{k}_i - \mathbf{k}_j \right\|^2 + \epsilon},
\vspace{-1.5pt}
\end{equation}
where $\epsilon$ is a small constant to ensure numerical stability. The overall training objective combines all components:
\begin{equation}
\vspace{-1.5pt}
\mathcal{L} = \mathcal{L}_{\text{recon}} + \eta \cdot \mathcal{L}_{\text{KL}} + \lambda \cdot \mathcal{L}_{\text{div}},
\vspace{-1.5pt}
\end{equation}
with hyperparameters $\eta$ and $\lambda$ controlling the importance of KL regularization and diversity.

After training, the generator produces a set of candidate keys for interest memory retrieval by sampling latent vectors from the standard Gaussian prior conditioned on the input key.

\subsubsection{Interest Retrieval}
Given the generated query keys, we retrieve relevant memory entries using Reciprocal Rank Fusion (RRF) mechanism, a rank-based ensemble technique that aggregates relevance scores from multiple queries. For a memory entry $m$, its RRF score is computed as:
\begin{equation}
\vspace{-1.5pt}
\mathrm{RRF}(m) = \sum_{\mathbf{k} \in \mathbf{K}} \frac{1}{\mathrm{rank}_{\mathbf{k},m} + a},
\vspace{-1pt}
\end{equation}
where $\mathrm{rank}_{\mathbf{k}, m}$ is the position of entry $m$ in the ranked list induced by cosine similarity between the query key $\mathbf{k}$ and the memory key of $m$. The constant $a$ is a smoothing factor to prevent score inflation for top-ranked entries.

The RRF scores are then normalized using a softmax function to compute attention weights for each memory entry. The sustained interest vector is obtained by aggregating the memory values weighted by these attention scores:
\begin{equation}
\vspace{-2pt}
\mathbf{I}_{\text{sustained}} = \sum_{m \in M_u} \mathrm{Softmax}\bigl(\mathrm{RRF}(m)\bigr) \cdot \mathbf{v}_m,
\end{equation}
where $\mathbf{v}_m$ denotes the value vector of memory entry $m$. The vector $\mathbf{I}_\text{sustained}$ reflects individual historical interests and is an important factor for producing final recommendations.

\subsection{Adaptive Interest Update and Fusion}
The adaptive interest update and fusion module is a critical component of GIRAM, serving distinct purposes in the update and deployment stages. During the update stage, it refines interest representations in response to evolving user preferences, while in the deployment stage, it integrates sustained and recent interests to generate recommendations. Despite their different purposes, the update process and fusion process are closely related to each other, and both require a nuanced combination of sustained interests and recent interests. 
% To this end, we propose employing consistency scores to adaptively guide both update and fusion processes, ensuring a balance between historical and recent information.
To this end, consistency scores are utilized to adaptively guide both update and fusion processes, thereby ensuring a balance between historical and recent information.

\subsubsection{Consistency Score}
\label{sec:consistency_score}
We introduce a consistency score to quantify the alignment between the outputs of the personalized model (with trainable user embeddings) and the collective model (with frozen user embeddings). Both models are finetuned on the new data block, but differ in whether the user embeddings are trainable or frozen. The personalized model captures individual preferences, while the collective model reflects population-level interest trends.

Intuitively, a high consistency score suggests a user's interests align with general trends, suggesting that recent contextual patterns should play a more important role. For example, if a user visits popular seasonal POIs (e.g., beaches or parks in summer), recommendations should emphasize trending activities. Conversely, a low consistency score indicates weak alignment, emphasizing personalized preferences. For instance, a running enthusiast may frequent parks in winter despite low popularity, requiring recommendations tailored to their unique interests rather than general trends. This mechanism dynamically balances historical preferences and current contextual trends in both memory updates and recommendations. 

The consistency score $s_u$ for user $u$ is computed as the cosine similarity between the output vectors of the two models:
\begin{equation}
\vspace{-2pt}
s_u = \frac{\mathbf{x}_c^u \cdot \mathbf{x}_p^u}{\|\mathbf{x}_c^u\| \, \|\mathbf{x}_p^u\|},
\vspace{-2pt}
\end{equation}
where $\mathbf{x}_c^u$ and $\mathbf{x}_p^u$ are the outputs of the collective and personalized models, respectively.

\subsubsection{Adaptive Interest Update}
For each new record, the interest memory update involves two steps. First, the update weight $\alpha_u$ is adjusted based on the consistency score:
\begin{equation}
\vspace{-2pt}
\alpha_u = \alpha_{\text{base}} + \gamma (s_u - s_{\text{mean}}),
\vspace{-2pt}
\end{equation}
where $\alpha_{\text{base}}$ is a predefined base value, $s_u$ is the user's consistency score, and $s_{\text{mean}}$ is the average consistency score across all users. The parameter $\gamma$ controls the sensitivity of the adjustment. This ensures that interactions with higher consistency scores trigger more significant updates, incorporating current interests, whereas lower scores result in less pronounced updates.

Second, we assess whether the incoming key matches an existing memory entry using cosine similarity function. If the maximum similarity between the incoming key and existing keys exceeds a predefined update, threshold $\delta$, the matched memory entry is updated as follows:
\begin{equation}
\vspace{-2pt}
\mathbf{k}_{\text{new}} = (1 - \alpha_u) \cdot \mathbf{k}_{\text{mtd}} + \alpha_u \cdot \mathbf{k}, 
\mathbf{v}_{\text{new}} = (1 - \alpha_u) \cdot \mathbf{v}_{\text{mtd}} + \alpha_u \cdot \mathbf{v},
\vspace{-2pt}
\end{equation}
where $\mathbf{k}_{\text{mtd}}$ and $\mathbf{v}_{\text{mtd}}$ are the matched key and value, and $\mathbf{k}$ and $\mathbf{v}$ are the incoming key and value. If the maximum similarity does not exceed $\delta$, a new memory entry is added if capacity allows; otherwise, the oldest entry is replaced. This ensures the memory remains dynamically updated and efficiently managed, reflecting evolving user preferences while maintaining capacity constraints.

\subsubsection{Adaptive Interest Fusion}
During deployment, the final interest representation is obtained by fusing recent and sustained interests, weighted by the user's consistency score. The fusion coefficient $\beta_u$ is computed as:
\begin{equation}
\vspace{-2pt}
\beta_u = \beta_{\text{base}} + \gamma(s_u - s_{\text{mean}}),
\vspace{-2pt}
\end{equation}
where $\beta_{\text{base}}$ is a predefined base weight. A higher $s_u$ increases the contribution of recent interests, while a lower $s_u$ emphasizes sustained interests. The final interest vector is:
\begin{equation}
\vspace{-2pt}
\mathbf{I} = (1- \beta_u) \cdot \mathbf{I}_{\text{sustained}} + \beta_u \cdot \mathbf{I}_{\text{recent}},
\vspace{-2pt}
\end{equation}
where $\mathbf{I}_{\text{recent}}$ is the output of the finetuned personalized model. This fusion strategy ensures that recommendations adapt to both persistent preferences and recent behavioral shifts.

\subsubsection{Procedures for Update and Deployment}
The pseudo-code for the update and deployment stages are presented in Algorithm~\ref{alg:update} and Algorithm~\ref{alg:deployment}, respectively.

During the update stage, a consistency score $s_u$ is computed for each user as the similarity between the outputs of $f_c$ and $f_p$ (Lines 1--4), followed by the computation of the mean consistency score $s_{\text{mean}}$ across users (Line 5). An adaptive update weight $\alpha_u$ is then derived based on $s_u$ and $s_{\text{mean}}$ (Line 7). For each check-in trajectory, a key-value pair is generated, and its similarity to existing memory entries is evaluated (Lines 9--10). If the similarity exceeds a predefined threshold $\delta$, the matched entry is updated using the weight $\alpha_u$ (Lines 11--14); otherwise, a new entry is inserted, or the oldest entry is replaced if the memory is full (Lines 15--18).

\begin{algorithm}[!t]
\caption{The Update Stage}\label{alg:update}
\begin{flushleft}
\textbf{Input}: Interest memory $M$, update threshold $\delta$,\\
\qquad\quad collective model $f_c$, personalized model $f_p$,\\
\qquad\quad training dataset $\mathcal{T}$, user set $\mathcal{U}$, memory capacity $N_u$,\\
\qquad\quad base update weight $\alpha_{\text{base}}$, sensitivity parameter $\gamma$\\
\textbf{Output}: Updated interest memory $\tilde{M}$
\end{flushleft}
\begin{algorithmic}[1]
\FOR{$u \in \mathcal{U}$}
    \STATE $\mathbf{x}_c^u \gets [f_c(T_u)]_{T_u \in \mathcal{T}_u}$, \quad $\mathbf{x}_p^u \gets [f_p(T_u)]_{T_u \in \mathcal{T}_u}$
    \STATE $s_u \gets \frac{\mathbf{x}_c^u \cdot \mathbf{x}_p^u}{\|\mathbf{x}_c^u\| \, \|\mathbf{x}_p^u\|}$
\ENDFOR
\STATE $s_{\text{mean}} \gets \frac{1}{|\mathcal{U}|} \sum_{u \in \mathcal{U}} s_u$
\FOR{$u \in \mathcal{U}$}
    \STATE $\alpha_u \gets \alpha_{\text{base}} + \gamma (s_u - s_{\text{mean}})$
    \FOR{$T_u \in \mathcal{T}_u$}
        \STATE $\mathbf{k} \gets \text{KeyEncoding}(T_u)$, $\mathbf{v} \gets f_p(T_u)$
        \STATE $(\text{mtd}, \text{max\_sim}) \gets \arg\max_{i \in M_u} \frac{\mathbf{k} \cdot \mathbf{k}_i}{\|\mathbf{k}\| \, \|\mathbf{k}_i\|}$
        \IF{$\textit{max\_sim} > \delta$}
            \STATE $\mathbf{k}_{\text{new}} \gets (1 - \alpha_u) \cdot \mathbf{k}_\text{mtd} + \alpha_u \cdot \mathbf{k}$
            \STATE $\mathbf{v}_{\text{new}} \gets (1 - \alpha_u) \cdot \mathbf{v}_\text{mtd} + \alpha_u \cdot \mathbf{v}$
            \STATE $M_u[\text{mtd}] \gets (\mathbf{k}_{\text{new}}, \mathbf{v}_{\text{new}}, t)$
        \ELSIF{$|M_u| < N_u$}
            \STATE $M_u \gets M_u \cup \{(\mathbf{k}, \mathbf{v}, t)\}$
        \ELSE
            \STATE $M_u[\text{FindOldestEntry}(M_u)] \gets (\mathbf{k}, \mathbf{v}, t)$
        \ENDIF
    \ENDFOR
\ENDFOR
\STATE \textbf{return} $\tilde{M} \gets M$
\end{algorithmic}
\end{algorithm}
\vspace{-1pt}
\begin{algorithm}[!t]
\caption{The Deployment Stage}\label{alg:deployment}
\begin{flushleft}
\textbf{Input}:
Interest memory $M_u$ of user $u$, personalized model $f_p$\\
\qquad\quad sensitivity parameter $\gamma$, consistency score $s_u$,\\
\qquad\quad average consistency score $s_{\text{mean}}$,\\
\qquad\quad a new collected trajectory $T_u$, base fusion weight $\beta_{\text{base}}$\\
\textbf{Output}: Final recommendation results $\mathbf{I}$
\end{flushleft}
\begin{algorithmic}[1] %[1] enables line numbers
\STATE $\mathbf{k} \gets \text{KeyEncoding}(T_u)$
\STATE $\mathbf{K} \gets [\mathbf{k}_{1}, \mathbf{k}_{2}, \cdots, \mathbf{k}_{N_k}]$
\FOR{$m \in M_u$}
    \STATE $\mathrm{RRF}(m) \gets \sum_{\mathbf{k} \in \mathbf{K}} \frac{1}{\text{rank}_{\mathbf{k},m} + a}$
\ENDFOR
\STATE $\mathbf{I}_\text{sustained} \gets \sum_{m\in M_u} \mathrm{Softmax}(\mathrm{RRF}(m)) \cdot \mathbf{v}_m$
\STATE $\mathbf{I}_\text{recent} \gets f_p(T_u)$
\STATE $\beta_u \gets \beta_{\text{base}} + \gamma(s_u - s_{\text{mean}})$;
\STATE $\mathbf{I} \gets (1- \beta_u) \cdot \mathbf{I}_{\text{sustained}} + \beta_u \cdot \mathbf{I}_{\text{recent}}$
\STATE \textbf{return} $\mathbf{I}$
\end{algorithmic}
\end{algorithm}
\vspace{-1pt}

During the deployment stage, context-aware keys $\mathbf{K}$ are first generated from the user's trajectory $T_u$ (Lines 1--2). Each memory entry in $M_u$ is then scored using Reciprocal Rank Fusion (RRF), based on its ranks across the generated keys (Lines 3--5). The sustained interest vector $\mathbf{I}_{\text{sustained}}$ is computed as a softmax-weighted sum of the memory values (Line 6), while the recent interest vector $\mathbf{I}_{\text{recent}}$ is obtained by applying the function $f_p$ to $T_u$ (Line 7). A fusion weight $\beta_u$ is then determined based on the user's relative consistency (Line 8), and the final recommendation vector $\mathbf{I}$ is produced by fusing the sustained and recent interests (Line 9).

\vspace{-2pt}
\subsection{Complexity Analysis}
\vspace{-2pt}
\label{sec:complexity}
\subsubsection{Update Time Complexity} The update process consists of computing consistency scores and updating the interest memory. Computing the consistency scores requires a time complexity of $\mathcal{O}(|\mathcal{T}| \cdot \bar{n} \cdot d_v)$, where $|\mathcal{T}|$ is the number of training trajectories, $\bar{n}$ is the average trajectory length, and $d_v$ is the dimensionality of the NPR model output. Updating the interest memory involves encoding $d_k$-dimensional keys from trajectories (average length $\bar{n}$) and matching them against $N_u$ memory entries per user, which incurs a time complexity of $\mathcal{O}\bigl(|\mathcal{T}| \cdot (\bar{n} + N_u) \cdot d_k\bigr)$.
Therefore, the overall time complexity of the update stage is $\mathcal{O}\Bigl(|\mathcal{T}| \cdot \bigl(\bar{n} \cdot d_v + (\bar{n} + N_u) \cdot d_k\bigr)\Bigr)$.

\subsubsection{Deployment Time Complexity} During deployment, a new trajectory is processed to generate $N_k$ context-aware keys, each of dimension $d_k$. These keys are then ranked and matched against $N_u$ memory entries to compute the RRF scores, which are subsequently used to generate sustained interest. Therefore, the total deployment time complexity of $\mathcal{O}\bigl(|\mathcal{T}| \cdot (\bar{n} \cdot N_k \cdot d_k + N_u \cdot N_k \cdot d_k + N_u \cdot d_v )\bigr)$.

\subsubsection{Space Complexity.} The main space cost arises from the interest memory. Each entry stores a $d_k$-dimensional key and a sparse value vector containing top-$K$ POI probabilities. Therefore, the memory complexity per-user is $\mathcal{O}\bigl(N_u \cdot (d_k + K)\bigr)$, and the total space complexity of all users is $\mathcal{O}\bigl(|U| \cdot N_u \cdot (d_k + K)\bigr)$.
\section{EXPERIMENTS}
\vspace{-2pt}
\subsection{Experimental Setup}\label{sec:5.1}
\vspace{-2pt}
\subsubsection{Datasets \& Preprocessing}
We evaluate on three real-world datasets: Foursquare-NYC~\cite{DBLP:journals/tsmc/YangZZY15}, Foursquare-TKY\footnote{\url{https://sites.google.com/site/yangdingqi/home/foursquare-dataset}}~\cite{DBLP:journals/tsmc/YangZZY15}, and Gowalla-CA\footnote{\url{https://snap.stanford.edu/data/loc-gowalla.html}}~\cite{DBLP:conf/kdd/ChoML11}, referred to as NYC, TKY, and CA, respectively. The datasets are preprocessed as follows: i) POIs and users with fewer than 10 check-in records are filtered out; ii) Check-in records are sorted chronologically, with the earliest 50\% check-in records forming the base data block $\mathcal{T}_0$, and the remaining check-ins uniformly divided into five blocks along time; iii) Within each block, user check-ins are divided into trajectories using one-week intervals, with trajectories containing only one check-in discarded, forming five incremental trajectory data blocks $[\mathcal{T}_1, \mathcal{T}_2, \mathcal{T}_3, \mathcal{T}_4, \mathcal{T}_5]$.

This setting reflects real-world practice, where models are updated periodically as new data is collected rather than continuously. In the long term, retaining the entire dataset is not feasible, but in the short term, the contextual information within each block is generally sufficient for model learning.

The model is initially trained on $\mathcal{T}_0$ and validated on the first incremental data block $\mathcal{T}_1$. For each subsequent step, the model is updated using the entire $\mathcal{T}_i$, and the next block $\mathcal{T}_{i+1}$ is randomly split in half for validation and testing. This update-and-evaluate process is repeated for all remaining blocks. Statistics of the processed datasets are presented in Table~\ref{tb:dataset}.

\renewcommand{\arraystretch}{0.8}
\setlength{\tabcolsep}{1mm}
\begin{table}[htp]
\centering{\fontsize{7.8}{9.6}\selectfont
\vspace{-12pt}
\caption{Dataset statistics.}
\vspace{-8pt}
\label{tb:dataset}
\begin{tabular}{lccccc}
\toprule
Dataset & \# Users & \# POIs & \# Trajectories & \# Check-ins & \# Collection Period\\
\midrule
NYC & 1,083 & 5,135 & 19,405 & 144,268 & 2012/04 – 2013/02 \\
TKY & 2,293 & 7,873 & 47,731 & 440,783 & 2012/04 – 2013/02 \\
CA & 6,592 & 14,027 & 48,697 & 338,612 & 2009/02 – 2010/10 \\
\bottomrule
\end{tabular}}
\vspace{-7pt}
\end{table}

\subsubsection{Backbone Models}
We apply our framework to three next POI recommendation models:
\begin{itemize}[leftmargin=*]
\item \textbf{Flashback}~\cite{DBLP:conf/ijcai/YangFRC20} is an RNN-based next POI recommender that leverages spatio-temporal context to retrieve historically relevant hidden states for improved accuracy.
\item \textbf{GETNext}~\cite{DBLP:conf/sigir/YangL022} is a method that uses trajectory flow maps to capture movement patterns. It introduces a graph-enhanced transformer to harness extensive collaborative signals.
\item \textbf{DiffPOI}~\cite{DBLP:journals/tois/QinWJLZ24} is a diffusion-based model that encodes visit sequences and spatial features with two graph encoders, using diffusion sampling to capture spatial visiting patterns.
\end{itemize}

We choose these backbones for their different architectures: Flashback is based on RNN, GETNext integrates GCN with Transformer, and DiffPOI combines GCN, CNN, and a diffusion model. This diversity allows a comprehensive evaluation of GIRAM using different structural backbones.

\subsubsection{Evaluation Metrics}
We adopt two widely used evaluation metrics: Top-$k$ Accuracy (Acc@$k$) and Mean Reciprocal Rank (MRR). Acc@$k$ measures whether the ground truth POI appears within the top-$k$ recommendations. In this paper, we report Acc@5, Acc@10 and Acc@20. MRR evaluates the position of the ground truth POI in the ranked recommendation list, assigning higher scores to predictions that are ranked higher. The two metrics are formally defined as follows:
\begin{equation}
\vspace{-2pt}
\text{Acc}@k = \frac{1}{N}\sum_{i=1}^{N}\mathbbm{1}(\text{rank}_i\le k),\ \text{MRR} = \frac{1}{N}\sum_{i=1}^{N}\frac{1}{\text{rank}_i},
\vspace{-2pt}
\end{equation}
where $N$ is the number of evaluation instances, $\mathbbm{1}(\cdot)$ is the indicator function, and $\text{rank}_i$ is the position of the ground truth POI for the $i$-th instance in the ranked recommendation list.

\subsubsection{Baselines}
We evaluate our method against the following strong and representative baselines:
\begin{itemize}[leftmargin=*]
\item \textbf{Retrain} trains a new model from scratch using all collected check-in data. The results from Retrain serve as a reference benchmark and are not included in the rankings due to its impracticality in continual learning settings.
\item \textbf{Static} uses a fixed model trained solely on the base data block, without incorporating any subsequent updates. It serves as a baseline in the absence of continual learning.
\item \textbf{Finetune} updates the model solely on the latest block, ignoring prior knowledge and risking catastrophic forgetting.
\item \textbf{ADER}~\cite{DBLP:conf/recsys/MiLF20} replays selected exemplars and applies adaptive knowledge distillation to preserve historical knowledge.
\item \textbf{IncCTR}~\cite{DBLP:journals/corr/abs-2009-02147} distills knowledge from the previous model by using its outputs as supervision for the current model.
\item \textbf{ReLoop2}~\cite{DBLP:conf/kdd/ZhuCHDT023} is a state-of-the-art replay-based continual learning method for recommendation, incorporating an error memory mechanism to estimate the discrepancy between predictions and ground-truth labels.
\item \textbf{CMuST}~\cite{DBLP:conf/nips/YiZHCYW024} is a spatio-temporal continual learning method combining an interaction network with rolling adaptation. We remove its task summarization module as it is tailored to continuous data and unsuitable for POI data.
\item \textbf{URCL}~\cite{DBLP:conf/icde/00010GY0HXJ24} is a continual spatio-temporal prediction framework combining replay buffer–based mixup, a spatio-temporal autoencoder, and STSimSiam. For CNPR, we remove STMixup and replace the encoder–decoder with POI recommendation backbones.
\end{itemize}

\begin{table*}[!t]
\vspace{-4pt}
\centering\caption{Performance comparison on NYC dataset.}\label{tb:NYC}
\vspace{-8pt}
{\fontsize{6.8}{7.6}\selectfont
\setlength{\tabcolsep}{0.16mm}
\renewcommand{\arraystretch}{1}
\begin{tabular}{c|l|cccc|cccc|cccc|cccc|cccc}
\hline
\multirow{2}*{Backbones} & \multirow{2}*{Methods} & \multicolumn{4}{c|}{$\mathcal{T}_2$} & \multicolumn{4}{c|}{$\mathcal{T}_3$} & \multicolumn{4}{c|}{$\mathcal{T}_4$} & \multicolumn{4}{c|}{$\mathcal{T}_5$} & \multicolumn{4}{c}{Mean}\\
\cline{3-22}
~& & Acc@5 & Acc@10 & Acc@20 & MRR & Acc@5 & Acc@10 & Acc@20 & MRR & Acc@5 & Acc@10 & Acc@20 & MRR & Acc@5 & Acc@10 & Acc@20 & MRR & Acc@5 & Acc@10 & Acc@20 & MRR\\
\hline
\multirow{9}*{Flashback} & Retrain & 0.5055 & 0.5716 & 0.6047 & 0.3614 & 0.4351 & 0.4826 & 0.5190 & 0.3113 & 0.4489 & 0.5210 & 0.5657 & 0.3070 & 0.5605 & 0.6351 & 0.6765 & 0.4014 & 0.4875 & 0.5526 & 0.5915 & 0.3453\\
\cline{2-22}
~& Static & 0.4506 & 0.4982 & 0.5408 & 0.3117 & 0.3530 & 0.4233 & 0.4573 & 0.2632 & 0.3352 & 0.3785 & 0.4146 & 0.2305 & 0.3330 & 0.4225 & 0.4598 & 0.2699 & 0.3680 & 0.4306 & 0.4681 & 0.2688\\
~& Finetune & 0.4879 & 0.5459 & 0.5900 & 0.3486 & 0.3995 & 0.4573 & 0.4929 & 0.2806 & 0.4361 & 0.4987 & 0.5348 & 0.3065 & \underline{0.5580} & 0.6181 & 0.6562 & \underline{0.4058} & 0.4704 & 0.5300 & 0.5685 & 0.3354\\
~& ADER & 0.4886 & 0.5511 & 0.5922 & 0.3515 & 0.4209 & 0.4747 & 0.5063 & 0.2918 & 0.4592 & \underline{0.5262} & 0.5657 & 0.3236 & 0.5360 & 0.5927 & 0.6207 & 0.3947 & 0.4762 & 0.5362 & 0.5712 & 0.3404\\
~& IncCTR & 0.4827 & 0.5334 & 0.5694 & 0.3450 & 0.4146 & 0.4794 & 0.5150 & 0.2874 & 0.4506 & 0.5202 & \underline{0.5665} & 0.3077 & 0.5326 & 0.6063 & 0.6427 & 0.3855 & 0.4701 & 0.5348 & 0.5734 & 0.3314\\
~& ReLoop2 & \underline{0.4960} & \underline{0.5709} & \underline{0.6054} & \underline{0.3607} & \underline{0.4399} & \underline{0.5047} & \textbf{0.5451} & \underline{0.3120} & 0.4232 & 0.5004 & 0.5468 & 0.3035 & 0.5326 & 0.6113 & 0.6630 & 0.3924 & 0.4729 & 0.5469 & \underline{0.5901} & \underline{0.3422}\\
~& CMuST & 0.4861 & 0.5483 & 0.5827 & 0.3386 & 0.4341 & 0.4966 & 0.5316 & 0.2927 & \textbf{0.4718} & 0.5250 & 0.5614 & \underline{0.3249} & 0.5477 & \underline{0.6340} & \underline{0.6672} & 0.3878 & \underline{0.4849} & \underline{0.5510} & 0.5857 & 0.3360\\
~& {URCL} & {0.4842} & {0.5511} & {0.5900} & {0.3376} & {0.4320} & {0.4921} & {0.5269} & {0.2965} & {0.4369} & {0.5013} & {0.5519} & {0.2975} & {0.5394} & {0.6164} & {0.6622} & {0.3798} & {0.4731} & {0.5402} & {0.5827} & {0.3278} \\
~& \textbf{GIRAM} & \textbf{0.5246} & \textbf{0.5907} & \textbf{0.6245} & \textbf{0.3773} & \textbf{0.4557} & \textbf{0.5127} & \underline{0.5443} & \textbf{0.3197} & \underline{0.4687} & \textbf{0.5373} & \textbf{0.5888} & \textbf{0.3338} & \textbf{0.5631} & \textbf{0.6410} & \textbf{0.6842} & \textbf{0.4131} & \textbf{0.5030} & \textbf{0.5704} & \textbf{0.6105} & \textbf{0.3610}\\
\hline
\multirow{9}*{GETNext} & Retrain & 0.4916 & 0.5650 & 0.6143 & 0.3459 & 0.4430 & 0.5150 & 0.5570 & 0.3074 & 0.4567 & 0.5348 & 0.6017 & 0.3061 & 0.5461 & 0.6410 & 0.6867 & 0.3854 & 0.4843 & 0.5640 & 0.6149 & 0.3362\\
\cline{2-22}
~& Static & 0.4445 & 0.5253 & 0.5650 & 0.3234 & 0.3877 & 0.4351 & 0.4620 & 0.2768 & 0.3270 & 0.3888 & 0.4335 & 0.2303 & 0.3421 & 0.3887 & 0.4335 & 0.2426 & 0.3753 & 0.4345 & 0.4735 & 0.2683\\
~& Finetune & 0.4570 & 0.5011 & 0.5298 & 0.3402 & 0.3979 & 0.4557 & 0.4889 & 0.2890 & \underline{0.4489} & \underline{0.5210} & 0.5562 & \underline{0.3278} & 0.5309 & 0.6088 & 0.6562 & 0.3888 & 0.4587 & 0.5217 & 0.5578 & 0.3364\\
~& ADER & 0.4783 & 0.5320 & 0.5687 & \underline{0.3530} & 0.4019 & 0.4644 & 0.5174 & 0.2900 & 0.4429 & 0.5030 & \underline{0.5597} & 0.3103 & 0.5445 & 0.6198 & 0.6638 & 0.3970 & 0.4669 & \underline{0.5298} & \underline{0.5774} & 0.3376\\
~& IncCTR & 0.4702 & 0.5217 & 0.5650 & 0.3399 & 0.4098 & 0.4620 & 0.5008 & 0.3027 & 0.4249 & 0.4901 & 0.5433 & 0.3051 & \textbf{0.5682} & \textbf{0.6384} & 0.6791 & \underline{0.4117} & \underline{0.4683} & 0.5281 & 0.5721 & 0.3398\\
~& ReLoop2 & 0.4702 & 0.5121 & 0.5489 & 0.3417 & 0.3956 & 0.4462 & 0.4810 & 0.2899 & 0.4343 & 0.4970 & 0.5356 & 0.3170 & 0.5631 & 0.6266 & \underline{0.6850} & \textbf{0.4168} & 0.4658 & 0.5205 & 0.5626 & \underline{0.3413}\\
~& CMuST & 0.4702 & 0.5143 & 0.5496 & 0.3470 & 0.3987 & 0.4533 & 0.4953 & 0.2811 & 0.4378 & 0.4901 & 0.5356 & 0.3171 & 0.5419 & 0.6190 & 0.6655 & 0.3998 & 0.4622 & 0.5192 & 0.5615 & 0.3363 \\
~& {URCL} & {\underline{0.4842}} & {\underline{0.5489}} & {\underline{0.5790}} & {\textbf{0.3532}} & {\underline{0.4146}} & {\underline{0.4786}} & {\underline{0.5277}} & {\underline{0.3064}} & {0.4249} & {0.4798} & {0.5296} & {0.3058} & {0.5267} & {0.5970} & {0.6486} & {0.3656} & {0.4626} & {0.5261} & {0.5712} & {0.3327} \\
~& \textbf{GIRAM} & \textbf{0.4871} & \textbf{0.5503} & \textbf{0.5907} & 0.3519 & \textbf{0.4343} & \textbf{0.4834} & \textbf{0.5269} & \textbf{0.3164} & \textbf{0.4644} & \textbf{0.5270} & \textbf{0.5785} & \textbf{0.3381} & \underline{0.5648} & \underline{0.6376} & \textbf{0.6867} & 0.4077 & \textbf{0.4877} & \textbf{0.5496} & \textbf{0.5957} & \textbf{0.3535}\\
\hline
\multirow{9}*{DiffPOI} & Retrain & 0.3865 & 0.4122 & 0.4372 & 0.3603 & 0.3608 & 0.3916 & 0.4177 & 0.3344 & 0.3888 & 0.4155 & 0.4438 & 0.3578 & 0.4920 & 0.5148 & 0.5487 & 0.4568 & 0.4070 & 0.4335 & 0.4618 & 0.3773\\
\cline{2-22}
~& Static & 0.3512 & 0.3747 & 0.4019 & 0.3273 & 0.3323 & 0.3568 & 0.3813 & 0.3020 & 0.3159 & 0.3468 & 0.3708 & 0.2849 & 0.3429 & 0.3683 & 0.3903 & 0.3086 & 0.3356 & 0.3617 & 0.3861 & 0.3057\\
~& Finetune & 0.3622 & 0.3916 & 0.4166 & 0.3313 & 0.3354 & 0.3592 & 0.3908 & 0.3117 & 0.3837 & \underline{0.4086} & \underline{0.4369} & 0.3510 & 0.4649 & 0.4928 & 0.5199 & 0.4277 & 0.3866 & 0.4130 & 0.4411 & 0.3554\\
~& ADER & \underline{0.3681} & 0.3902 & 0.4144 & \underline{0.3447} & 0.3489 & 0.3805 & 0.4074 & 0.3177 & 0.3837 & 0.4052 & 0.4275 & \underline{0.3532} & 0.4793 & 0.4987 & 0.5216 & \underline{0.4512} & 0.3950 & 0.4186 & 0.4427 & \underline{0.3667}\\
~& IncCTR & 0.3622 & \underline{0.3968} & \underline{0.4203} & 0.3338 & \underline{0.3703} & \underline{0.3916} & \underline{0.4375} & \underline{0.3316} & 0.3794 & 0.4034 & 0.4309 & 0.3500 & \underline{0.4860} & \underline{0.5080} & \underline{0.5318} & 0.4456 & \underline{0.3995} & \underline{0.4250} & \underline{0.4551} & 0.3652 \\
~& ReLoop2 & 0.3652 & 0.3916 & \underline{0.4203} & 0.3380 & 0.3378 & 0.3679 & 0.4059 & 0.3137 & \underline{0.3880} & 0.4034 & 0.4343 & 0.3526 & 0.4784 & 0.5021 & 0.5233 & 0.4430 & 0.3923 & 0.4163 & 0.4459 & 0.3618 \\
~& CMuST & 0.3581 & 0.3853 & 0.4096 & 0.3300 & 0.3259 & 0.3647 & 0.4051 & 0.3033 & 0.3674 & 0.4026 & 0.4361 & 0.3435 & 0.4644 & 0.4915 & 0.5271 & 0.4339 & 0.3790 & 0.4110 & 0.4444 & 0.3527 \\
~& {URCL} & {0.3542} & {0.3777} & {0.4041} & {0.3285} & {0.3457} & {0.3726} & {0.4138} & {0.3182} & {0.3639} & {0.3906} & {0.4180} & {0.3331} & {0.4496} & {0.4784} & {0.5030} & {0.4214} & {0.3784} & {0.4048} & {0.4347} & {0.3503} \\
~& \textbf{GIRAM} & \textbf{0.5276} & \textbf{0.5636} & \textbf{0.5915} & \textbf{0.4960} & \textbf{0.4581} & \textbf{0.4889} & \textbf{0.5206} & \textbf{0.4143} & \textbf{0.4850} & \textbf{0.5142} & \textbf{0.5485} & \textbf{0.4441} & \textbf{0.5724} & \textbf{0.6003} & \textbf{0.6308} & \textbf{0.5242} & \textbf{0.5107} & \textbf{0.5417} & \textbf{0.5728} & \textbf{0.4697}\\
\hline
\end{tabular}}
\vspace{-10pt}
\end{table*}

\begin{table*}[!ht]
\setlength{\tabcolsep}{0.16mm}
\centering
\caption{Performance comparison on TKY dataset.}\vspace{-8pt}\label{tb:TKY}
{\fontsize{6.8}{7.6}\selectfont
\setlength{\tabcolsep}{0.16mm}
\renewcommand{\arraystretch}{1}
\begin{tabular}{c|l|cccc|cccc|cccc|cccc|cccc}
\hline
\multirow{2}*{Backbones} & \multirow{2}*{Methods} & \multicolumn{4}{c|}{$\mathcal{T}_2$} & \multicolumn{4}{c|}{$\mathcal{T}_3$} & \multicolumn{4}{c|}{$\mathcal{T}_4$} & \multicolumn{4}{c|}{$\mathcal{T}_5$} & \multicolumn{4}{c}{Mean}\\
\cline{3-22}
~& & Acc@5 & Acc@10 & Acc@20 & MRR & Acc@5 & Acc@10 & Acc@20 & MRR & Acc@5 & Acc@10 & Acc@20 & MRR & Acc@5 & Acc@10 & Acc@20 & MRR & Acc@5 & Acc@10 & Acc@20 & MRR\\
\hline
\multirow{9}*{Flashback} & Retrain & 0.4413 & 0.5269 & 0.5951 & 0.3177 & 0.4344 & 0.5201 & 0.5919 & 0.3191 & 0.4083 & 0.4933 & 0.5629 & 0.2902 & 0.4745 & 0.5496 & 0.6232 & 0.3308 & 0.4396 & 0.5225 & 0.5932 & 0.3144\\
\cline{2-22}
~& Static & 0.3795 & 0.4814 & 0.5402 & 0.2873 & 0.3570 & 0.4528 & 0.5118 & 0.2668 & 0.3158 & 0.3908 & 0.4502 & 0.2223 & 0.3245 & 0.4149 & 0.4667 & 0.2471 & 0.3442 & 0.4350 & 0.4922 & 0.2559\\
~& Finetune & 0.4038 & 0.4705 & 0.5269 & 0.2876 & 0.4364 & 0.5090 & 0.5656 & 0.3109 & 0.4014 & 0.4812 & 0.5450 & 0.2885 & 0.4667 & 0.5484 & 0.6065 & 0.3347 & 0.4271 & 0.5023 & 0.5610 & 0.3054\\
~& ADER & 0.4038 & 0.4655 & 0.5216 & 0.2812 & 0.4233 & 0.5010 & 0.5723 & 0.3114 & 0.4014 & 0.4798 & 0.5428 & 0.2882 & 0.4427 & 0.5311 & 0.5858 & 0.3194 & 0.4178 & 0.4943 & 0.5556 & 0.3001\\
~& IncCTR & 0.4064 & 0.4739 & 0.5292 & 0.2894 & 0.4293 & 0.5022 & 0.5660 & 0.3131 & 0.3999 & 0.4874 & 0.5530 & 0.2891 & 0.4675 & 0.5418 & 0.6128 & 0.3386 & 0.4258 & 0.5013 & 0.5652 & 0.3075 \\
~& ReLoop2 & \underline{0.4398} & \underline{0.5102} & \underline{0.5705} & \textbf{0.3371} & \underline{0.4512} & \underline{0.5193} & \underline{0.5946} & \underline{0.3201} & \underline{0.4316} & \underline{0.5159} & \underline{0.5821} & \underline{0.3126} & \underline{0.4689} & \underline{0.5603} & \underline{0.6257} & \underline{0.3499} & \underline{0.4479} & \underline{0.5264} & \underline{0.5932} & \underline{0.3299}\\
~& CMuST & 0.4042 & 0.4705 & 0.5303 & 0.2932 & 0.4352 & 0.5078 & 0.5688 & 0.3124 & 0.4094 & 0.4801 & 0.5461 & 0.2937 & 0.4663 & 0.5503 & 0.6202 & 0.3418 & 0.4288 & 0.5022 & 0.5663 & 0.3103 \\
~& {URCL} & {0.4220} & {0.4996} & {0.5602} & {0.3079} & {0.3986} & {0.4735} & {0.5460} & {0.2906} & {0.3828} & {0.4630} & {0.5341} & {0.2818} & {0.4508} & {0.5392} & {0.6102} & {0.3264} & {0.4135} & {0.4938} & {0.5626} & {0.3017} \\
~& \textbf{GIRAM} & \textbf{0.4508} & \textbf{0.5239} & \textbf{0.5795} & \underline{0.3270} & \textbf{0.4528} & \textbf{0.5269} & \textbf{0.5986} & \textbf{0.3230} & \textbf{0.4375} & \textbf{0.5210} & \textbf{0.5866} & \textbf{0.3142} & \textbf{0.5063} & \textbf{0.5954} & \textbf{0.6461} & \textbf{0.3683} & \textbf{0.4618} & \textbf{0.5418} & \textbf{0.6027} & \textbf{0.3331}\\
\hline
\multirow{9}*{GETNext} & Retrain & 0.4348 & 0.5102 & 0.5761 & 0.3182 & 0.4444 & 0.5341 & 0.6014 & 0.3163 & 0.4160 & 0.5031 & 0.5684 & 0.2937 & 0.4822 & 0.5762 & 0.6376 & 0.3423 & 0.4444 & 0.5309 & 0.5959 & 0.3176\\
\cline{2-22}
~& Static & 0.4148 & {0.4879} & {0.5455} & 0.3029 & 0.3627 & 0.4316 & 0.4922 & 0.2604 & 0.3226 & 0.3923 & 0.4561 & 0.2294 & 0.3502 & 0.4146 & 0.4689 & 0.2486 & 0.3626 & 0.4316 & 0.4907 & 0.2603\\
~& Finetune & 0.3587 & 0.4318 & 0.4898 & 0.2621 & 0.4101 & 0.4835 & 0.5520 & 0.2942 & 0.4200 & 0.4965 & 0.5578 & 0.3027 & \textbf{0.4723} & 0.5447 & 0.6080 & 0.3335 & 0.4153 & 0.4891 & 0.5519 & 0.2981\\
~& ADER & \underline{0.4239} & 0.4838 & 0.5433 & \underline{0.3077} & 0.4392 & 0.5189 & 0.5699 & 0.3142 & 0.4138 & 0.5016 & 0.5575 & 0.3032 & 0.4634 & 0.5440 & 0.6043 & 0.3348 & \underline{0.4351} & 0.5121 & 0.5687 & 0.3150\\
~& IncCTR & 0.4057 & 0.4784 & 0.5451 & 0.2984 & \underline{0.4508} & \underline{0.5229} & \underline{0.5851} & 0.3079 & 0.4229 & 0.5027 & \textbf{0.5673} & 0.2988 & 0.4601 & \underline{0.5510} & 0.6050 & 0.3300 & 0.4349 & \underline{0.5138} & \underline{0.5756} & 0.3088\\
~& ReLoop2 & 0.3875 & 0.4511 & 0.5023 & 0.2942 & 0.4408 & 0.5126 & 0.5747 & \underline{0.3223} & 0.4225 & 0.4936 & 0.5585 & \textbf{0.3090} & 0.4638 & 0.5418 & \underline{0.6106} & \underline{0.3403} & 0.4287 & 0.4998 & 0.5615 & \underline{0.3165}\\
~& CMuST & 0.3860 & 0.4481 & 0.4989 & 0.2764 & 0.4312 & 0.5082 & 0.5616 & 0.3072 & \underline{0.4269} & \textbf{0.5071} & 0.5618 & \underline{0.3035} & 0.4375 & 0.5141 & 0.5839 & 0.3174 & 0.4204 & 0.4944 & 0.5515 & 0.3011 \\
& {URCL} & {0.4140} & {\underline{0.4886}} & {\textbf{0.5530}} & {0.2934} & {0.4237} & {0.5098} & {0.5684} & {0.3054} & {0.3974} & {0.4736} & {0.5359} & {0.2866} & {0.4660} & {0.5570} & {0.6199} & {0.3329} & {0.4253} & {0.5072} & {0.5693} & {0.3046} \\
~& \textbf{GIRAM} & \textbf{0.4292} & \textbf{0.4951} & \underline{0.5462} & \textbf{0.3127} & \textbf{0.4560} & \textbf{0.5265} & \textbf{0.5859} & \textbf{0.3330} & \textbf{0.4273} & \underline{0.5057} & \underline{0.5662} & 0.3031 & \underline{0.4693} & \textbf{0.5573} & \textbf{0.6195} & \textbf{0.3446} & \textbf{0.4454} & \textbf{0.5211} & \textbf{0.5794} & \textbf{0.3234}\\
\hline
\multirow{9}*{DiffPOI} & Retrain & 0.4098 & 0.4580 & 0.5030 & 0.3588 & 0.4225 & 0.4631 & 0.5122 & 0.3708 & 0.4076 & 0.4521 & 0.4984 & 0.3506 & 0.4593 & 0.5033 & 0.5492 & 0.4086 & 0.4248 & 0.4691 & 0.5157 & 0.3722 \\
\cline{2-22}
~& Static & 0.4110 & 0.4542 & 0.5000 & 0.3604 & 0.3655 & 0.4117 & 0.4611 & 0.3151 & 0.3558 & 0.4036 & 0.4389 & 0.3010 & 0.3802 & 0.4231 & 0.4686 & 0.3321 & 0.3781 & 0.4231 & 0.4672 & 0.3272\\
~& Finetune & 0.4008 & 0.4519 & 0.4996 & 0.3523 & 0.4161 & 0.4544 & 0.5026 & 0.3676 & 0.3897 & 0.4360 & 0.4794 & 0.3455 & 0.4430 & 0.4745 & 0.5207 & 0.3971 & 0.4124 & 0.4542 & 0.5006 & 0.3657 \\
~& ADER & 0.4170 & 0.4587 & 0.4970 & 0.3683 & 0.4165 & 0.4583 & 0.5014 & 0.3674 & \underline{0.4028} & \underline{0.4419} & 0.4878 & \underline{0.3473} & \underline{0.4638} & \underline{0.5048} & \underline{0.5477} & \underline{0.4151} & \underline{0.4250} & \underline{0.4659} & 0.5085 & \underline{0.3745} \\
~& IncCTR & \underline{0.4189} & 0.4583 & 0.5061 & 0.3626 & 0.4109 & \underline{0.4587} & \underline{0.5070} & 0.3634 & 0.3966 & 0.4393 & \underline{0.4911} & 0.3451 & 0.4589 & 0.4985 & 0.5381 & 0.4096 & 0.4214 & 0.4637 & \underline{0.5105} & 0.3702 \\
~& ReLoop2 & 0.4167 & \underline{0.4670} & \underline{0.5170} & \underline{0.3720} & 0.4149 & 0.4552 & 0.5006 & \underline{0.3709} & 0.3930 & 0.4397 & 0.4794 & 0.3463 & 0.4464 & 0.4819 & 0.5277 & 0.3970 & 0.4177 & 0.4609 & 0.5062 & 0.3715 \\
~& CMuST & 0.4095 & 0.4527 & 0.4955 & 0.3596 & \underline{0.4195} & 0.4533 & 0.4952 & 0.3707 & 0.3913 & 0.4329 & 0.4796 & 0.3423 & 0.4462 & 0.4865 & 0.5276 & 0.3950 & 0.4166 & 0.4563 & 0.4995 & 0.3669 \\
~& {URCL} & {0.4064} & {0.4473} & {0.4913} & {0.3609} & {0.4014} & {0.4468} & {0.4898} & {0.3595} & {0.3996} & {0.4415} & {0.4845} & {0.3456} & {0.4438} & {0.4841} & {0.5300} & {0.3930} & {0.4128} & {0.4549} & {0.4989} & {0.3648} \\
~& \textbf{GIRAM} & \textbf{0.4833} & \textbf{0.5284} & \textbf{0.5720} & \textbf{0.4382} & \textbf{0.5114} & \textbf{0.5588} & \textbf{0.5943} & \textbf{0.4567} & \textbf{0.4860} & \textbf{0.5297} & \textbf{0.5731} & \textbf{0.4358} & \textbf{0.5314} & \textbf{0.5684} & \textbf{0.6135} & \textbf{0.4838} & \textbf{0.5030} & \textbf{0.5463} & \textbf{0.5882} & \textbf{0.4536} \\
\hline
\end{tabular}}
\vspace{-15pt}
\end{table*}

\begin{table*}[!t]\scriptsize
\vspace{-4pt}
\centering
\caption{Performance comparison on CA dataset.}\vspace{-8pt}\label{tb:CA}
{\fontsize{6.8}{7.6}\selectfont
\setlength{\tabcolsep}{0.16mm}
\renewcommand{\arraystretch}{1}
\begin{tabular}{c|l|cccc|cccc|cccc|cccc|cccc}
\hline
\multirow{2}*{Backbones} & \multirow{2}*{Methods} & \multicolumn{4}{c|}{$\mathcal{T}_2$} & \multicolumn{4}{c|}{$\mathcal{T}_3$} & \multicolumn{4}{c|}{$\mathcal{T}_4$} & \multicolumn{4}{c|}{$\mathcal{T}_5$} & \multicolumn{4}{c}{Mean}\\
\cline{3-22}
~& & Acc@5 & Acc@10 & Acc@20 & MRR & Acc@5 & Acc@10 & Acc@20 & MRR & Acc@5 & Acc@10 & Acc@20 & MRR & Acc@5 & Acc@10 & Acc@20 & MRR & Acc@5 & Acc@10 & Acc@20 & MRR\\
\hline
\multirow{9}*{Flashback} & Retrain & 0.2435 & 0.2913 & 0.3432 & 0.1821 & 0.2536 & 0.3053 & 0.3508 & 0.1881 & 0.2579 & 0.3232 & 0.3640 & 0.1918 & 0.3039 & 0.3574 & 0.4027 & 0.2180 & 0.2647 & 0.3193 & 0.3652 & 0.1950\\
\cline{2-22}
~& Static & 0.1876 & 0.2362 & 0.2796 & 0.1343 & 0.1518 & 0.1990 & 0.2344 & 0.1111 & 0.1302 & 0.1717 & 0.2097 & 0.0988 & 0.1164 & 0.1573 & 0.1908 & 0.0881 & 0.1465 & 0.1911 & 0.2286 & 0.1081\\
~& Finetune & 0.2306 & 0.2816 & 0.3274 & 0.1720 & 0.2328 & 0.2869 & 0.3256 & 0.1712 & 0.2460 & 0.3009 & 0.3467 & 0.1798 & 0.2798 & 0.3349 & 0.3777 & 0.2081 & 0.2473 & 0.3011 & 0.3443 & 0.1828\\
~& ADER & 0.2370 & 0.2844 & 0.3286 & 0.1761 & 0.2275 & 0.2825 & 0.329 & 0.1665 & 0.2509 & 0.2988 & 0.3520 & 0.1769 & 0.2777 & 0.3251 & 0.3730 & 0.2019 & 0.2483 & 0.2977 & 0.3458 & 0.1804\\
~& IncCTR & 0.2399 & 0.2893 & 0.3323 & 0.1762 & 0.2352 & 0.2865 & 0.3317 & 0.1743 & 0.2526 & 0.3054 & 0.3615 & \underline{0.1877} & \underline{0.2916} & 0.3412 & 0.3930 & 0.2130 & \underline{0.2548} & 0.3056 & 0.3546 & 0.1878\\
~& ReLoop2 & 0.2427 & \underline{0.3051} & \underline{0.3558} & \underline{0.1796} & 0.2304 & 0.2849 & 0.3337 & 0.1742 & \underline{0.2530} & \underline{0.3108} & \underline{0.3723} & 0.1830 & 0.2878 & \underline{0.3434} & \underline{0.3981} & \underline{0.2200} & 0.2535 & \underline{0.3110} & \underline{0.3650} & \underline{0.1892}\\
~& CMuST & \underline{0.2455} & 0.2929 & 0.3355 & 0.1766 & \underline{0.2438} & \underline{0.2959} & \underline{0.3390} & \underline{0.1795} & 0.2485 & 0.3000 & 0.3562 & 0.1773 & 0.2789 & 0.3319 & 0.3841 & 0.1978 & 0.2542 & 0.3052 & 0.3537 & 0.1828 \\
~& {URCL} & {0.2322} & {0.2869} & {0.3298} & {0.1707} & {0.2316} & {0.2808} & {0.3362} & {0.1660} & {0.2303} & {0.2848} & {0.3372} & {0.1675} & {0.2649} & {0.3247} & {0.3942} & {0.1957} & {0.2397} & {0.2943} & {0.3494} & {0.1750} \\
~& \textbf{GIRAM} & \textbf{0.2569} & \textbf{0.3169} & \textbf{0.3643} & \textbf{0.1907} & \textbf{0.2629} & \textbf{0.3109} & \textbf{0.3635} & \textbf{0.1957} & \textbf{0.2769} & \textbf{0.3322} & \textbf{0.3801} & \textbf{0.1983} & \textbf{0.3124} & \textbf{0.3709} & \textbf{0.4192} & \textbf{0.2267} & \textbf{0.2773} & \textbf{0.3327} & \textbf{0.3818} & \textbf{0.2028}\\
\hline
\multirow{9}*{GETNext} & Retrain & 0.2378 & 0.3015 & 0.3639 & 0.1733 & 0.2247 & 0.2869 & 0.3455 & 0.1647 & 0.2480 & 0.3132 & 0.3706 & 0.1805 & 0.2836 & 0.3548 & 0.4163 & 0.2081 & 0.2485 & 0.3141 & 0.3741 & 0.1816\\
\cline{2-22}
~& Static & 0.1746 & 0.2212 & 0.2694 & 0.1297 & 0.1457 & 0.1917 & 0.2324 & 0.1047 & 0.1226 & 0.1593 & 0.1948 & 0.0878 & 0.1162 & 0.1501 & 0.1895 & 0.0857 & 0.1398 & 0.1806 & 0.2215 & 0.1020\\
~& Finetune & 0.2241 & 0.2711 & 0.3177 & 0.1599 & 0.2120 & 0.2593 & 0.3089 & 0.1550 & 0.2501 & 0.3029 & 0.3516 & 0.1771 & \underline{0.2912} & 0.3480 & 0.4057 & 0.2136 & 0.2444 & 0.2953 & 0.3460 & 0.1764 \\
~& ADER & \underline{0.2366} & \underline{0.2869} & 0.3367 & 0.1668 & 0.2271 & 0.2808 & 0.3297 & 0.1642 & 0.2468 & 0.2984 & 0.3496 & 0.1795 & 0.2815 & 0.3480 & \underline{0.4078} & 0.2011 & 0.2480 & \underline{0.3035} & \underline{0.3559} & 0.1779 \\
~& IncCTR & 0.2257 & 0.2828 & \underline{0.3371} & 0.1604 & \underline{0.2304} & \underline{0.2841} & \underline{0.3374} & \underline{0.1648} & 0.2414 & 0.2988 & 0.3512 & 0.1676 & 0.2692 & 0.3302 & 0.3811 & 0.1977 & 0.2417 & 0.2990 & 0.3517 & 0.1726\\
~& ReLoop2 & 0.2285 & 0.2703 & 0.3160 & 0.1645 & 0.2161 & 0.2580 & 0.3118 & 0.1631 & \underline{0.2592} & \underline{0.3083} & \underline{0.3574} & \textbf{0.1890} & \textbf{0.3027} & \textbf{0.3544} & 0.4010 & \textbf{0.2271} & \underline{0.2516} & 0.2977 & 0.3466 & \underline{0.1859}\\
~& CMuST & 0.2314 & 0.2848 & 0.3318 & \underline{0.1687} & 0.2177 & 0.2690 & 0.3179 & 0.1606 & 0.2447 & 0.3042 & 0.3496 & 0.1705 & 0.2866 & 0.3442 & 0.3925 & 0.2129 & 0.2451 & 0.3006 & 0.3480 & 0.1782 \\
~& {URCL} & {0.2310} & {0.2751} & {0.3250} & {0.1649} & {0.2206} & {0.2747} & {0.3280} & {0.1587} & {0.2303} & {0.2827} & {0.3368} & {0.1588} & {0.2738} & {0.3298} & {0.3832} & {0.1966} & {0.2389} & {0.2906} & {0.3432} & {0.1697} \\
~& \textbf{GIRAM} & \textbf{0.2472} & \textbf{0.3011} & \textbf{0.3464} & \textbf{0.1829} & \textbf{0.2361} & \textbf{0.2918} & \textbf{0.3431} & \textbf{0.1765} & \textbf{0.2629} & \textbf{0.3157} & \textbf{0.3669} & \underline{0.1884} & 0.2895 & \underline{0.3518} & \textbf{0.4133} & \underline{0.2203} & \textbf{0.2589} & \textbf{0.3151} & \textbf{0.3674} & \textbf{0.1920} \\
\hline
\multirow{9}*{DiffPOI} & Retrain & 0.2338 & 0.2658 & 0.3071 & 0.1994 & 0.2389 & 0.2780 & 0.3171 & 0.2036 & 0.2526 & 0.2959 & 0.3401 & 0.2159 & 0.2857 & 0.3268 & 0.3777 & 0.2404 & 0.2527 & 0.2916 & 0.3355 & 0.2148 \\
\cline{2-22}
~& Static & 0.1892 & 0.2229 & 0.2581 & 0.1557 & 0.1734 & 0.2039 & 0.2377 & 0.1396 & 0.1721 & 0.2022 & 0.2410 & 0.1426 & 0.1624 & 0.2001 & 0.2336 & 0.1297 & 0.1743 & 0.2073 & 0.2426 & 0.1419\\
~& Finetune & 0.2107 & 0.2415 & 0.2848 & 0.1807 & 0.2145 & 0.2527 & 0.2906 & 0.1820 & 0.2344 & 0.2753 & 0.3120 & 0.2040 & 0.2675 & 0.3065 & 0.3463 & 0.2355 & 0.2318 & 0.2690 & 0.3084 & 0.2005 \\
~& ADER & \underline{0.2382} & \underline{0.2699} & \underline{0.3047} & \underline{0.2013} & \underline{0.2324} & \underline{0.2650} & 0.3073 & \underline{0.2011} & \underline{0.2575} & \underline{0.2914} & \underline{0.3306} & \underline{0.2200} & 0.2641 & 0.3069 & 0.3497 & 0.2333 & \underline{0.2481} & \underline{0.2833} & \underline{0.3231} & \underline{0.2139} \\
~& IncCTR & 0.2172 & 0.2496 & 0.2893 & 0.1827 & 0.2230 & 0.2621 & \underline{0.3093} & 0.1901 & 0.2402 & 0.2753 & 0.3219 & 0.2066 & \underline{0.2789} & \underline{0.3158} & \underline{0.3578} & \underline{0.2370} & 0.2398 & 0.2757 & 0.3196 & 0.2041 \\
~& ReLoop2 & 0.2131 & 0.2492 & 0.2869 & 0.1787 & 0.2226 & 0.2564 & 0.3040 & 0.1878 & 0.2381 & 0.2786 & 0.3170 & 0.2017 & 0.2683 & 0.3082 & 0.3455 & 0.2313 & 0.2356 & 0.2731 & 0.3133 & 0.1999 \\
~& CMuST & 0.2139 & 0.2488 & 0.2840 & 0.1795 & 0.1991 & 0.2370 & 0.2846 & 0.1671 & 0.2391 & 0.2709 & 0.3113 & 0.2062 & 0.2662 & 0.2980 & 0.3446 & 0.2277 & 0.2296 & 0.2637 & 0.3061 & 0.1952 \\
~& {URCL} & {0.2196} & {0.2549} & {0.2865} & {0.1805} & {0.2112} & {0.2491} & {0.2837} & {0.1811} & {0.2319} & {0.2650} & {0.3066} & {0.2008} & {0.2594} & {0.2900} & {0.3370} & {0.2297} & {0.2306} & {0.2647} & {0.3034} & {0.1980} \\
~& \textbf{GIRAM} & \textbf{0.2812} & \textbf{0.3189} & \textbf{0.3590} & \textbf{0.2432} & \textbf{0.2727} & \textbf{0.3081} & \textbf{0.3626} & \textbf{0.2331} & \textbf{0.2823} & \textbf{0.3190} & \textbf{0.3698} & \textbf{0.2456} & \textbf{0.3167} & \textbf{0.3527} & \textbf{0.3955} & \textbf{0.2750} & \textbf{0.2882} & \textbf{0.3247} & \textbf{0.3717} & \textbf{0.2492} \\
\hline
\end{tabular}}
\vspace{-12pt}
\end{table*}

\subsubsection{Implementation Details}
\label{sec:implementaion}
We implement GIRAM using Python 3.10 and PyTorch 2.1, and run all experiments on an NVIDIA GeForce RTX 4090 GPU. The interest memory capacity $N_u$ is set to 100 for NYC and TKY, and 20 for CA. We set the Top-$K$ value for memory construction to 50. All geographic, temporal, and categorical embeddings have dimension 16, and the latent dimension of key generator MLPs is 128. The sensitivity parameter $\gamma$ is set to 0.5. We set the KL divergence weight $\eta$ to 1, and the diversity loss weight $\lambda$ to 0.1. The smoothing factor $a$ is set to 50. In our experiments, the base update weight and fusion weight are set to 0.5. The number of generated keys $N_k$ is set to 20. The similarity threshold $\delta$ for memory updates is set to 0.95. For the backbones, we use the official implementation and keep the original hyperparameter configuration. The corresponding repositories are: Flashback\footnote{\url{https://github.com/eXascaleInfolab/Flashback_code}}, GETNext\footnote{\url{https://github.com/songyangme/GETNext}}, and DiffPOI\footnote{\url{https://github.com/Yifang-Qin/Diff-POI}}.

\vspace{-5pt}
\subsection{Main Results}
\vspace{-5pt}
The performance comparisons across the three datasets are shown in Table~\ref{tb:NYC}, Table~\ref{tb:TKY}, and Table~\ref{tb:CA}, respectively. The results are reported for each data block $[\mathcal{T}_2, \mathcal{T}_3, \mathcal{T}_4, \mathcal{T}_5]$ as well as the average metrics across all blocks. We can make the following observations: (1) Static underperforms across all metrics and datasets, as it is trained only on the base block without updates, leading to steadily declining performance as user behavior evolves. (2) Finetune shows inferior performance among updating methods, since it trains only on new data while neglecting historical information, causing catastrophic forgetting. (3) ADER outperforms Finetune by using frequency-based sample selection and replay to retain important historical check-ins. Its gain is especially notable with DiffPOI, whose stable diffusion architecture favors POIs present in historical interactions, reinforced by frequency-based selection. (4) IncCTR balances stability and adaptability by distilling knowledge from previous models while incorporating new data. However, its reliance on past outputs limits flexibility when new data diverge from historical patterns, a weakness most evident on the sparse and volatile CA dataset. (5) ReLoop2 is competitive with Flashback as the backbone, as its error memory mechanism enables targeted replay of hard examples, improving adaptation to temporal shifts. Yet its performance drops with GETNext and DiffPOI, showing sensitivity of the replay mechanism to backbone choice. (6) CMuST surpasses Finetune via weight behavior modeling, which tracks and adapts to gradual user-pattern changes. Nonetheless, its design for continuous prediction (e.g., traffic forecasting) constrains performance in discrete and sparse tasks like next POI recommendation. {(7) URCL performs well with GETNext, as its data augmentation aligns with spatio-temporal trajectory flow graphs. However, its overall effectiveness is limited, being tailored to continuous data (e.g., traffic flow) rather than discrete POI data.} (8) GIRAM, our proposed framework, consistently outperforms all baselines across datasets and backbone models. This superior performance validates the effectiveness of its architecture in capturing both sustained and recent user interests in a model-agnostic manner.

{Notably, GIRAM surpasses all baselines and often outperforms Retrain, underscoring its strong continual recommendation capability. Unlike Retrain, which may suffer from noise, outdated behaviors, or misaligned patterns in historical data, GIRAM selectively preserves and exploits relevant information, reducing interference while enhancing effectiveness and efficiency. These benefits are most evident with DiffPOI, whose assumption of a single check-in sequence per user limits its ability to capture diverse behaviors even with full retraining. GIRAM overcomes this limitation through a context-aware interest memory that leverages user-specific spatio-temporal information, yielding consistently superior performance.}

\subsection{Efficiency Study}
\subsubsection{Updating Time}
\begin{figure*}[!t]
\centering
\subfigure[{Flashback.}] {
\begin{minipage}{0.28\linewidth}
\centering
\vspace{-6pt}
\includegraphics[width=1\linewidth]{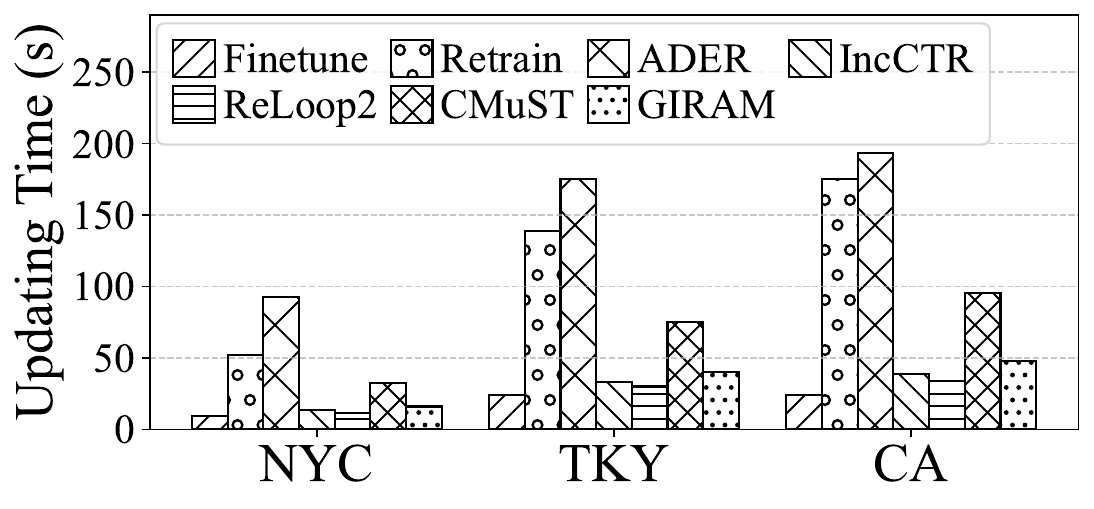}
\vspace{-10pt} 
\label{fig:time_Flashback}
\end{minipage}}
\subfigure[{GETNext.}]{
\begin{minipage}{0.28\linewidth}
\centering
\vspace{-2pt}
\includegraphics[width=1\linewidth]{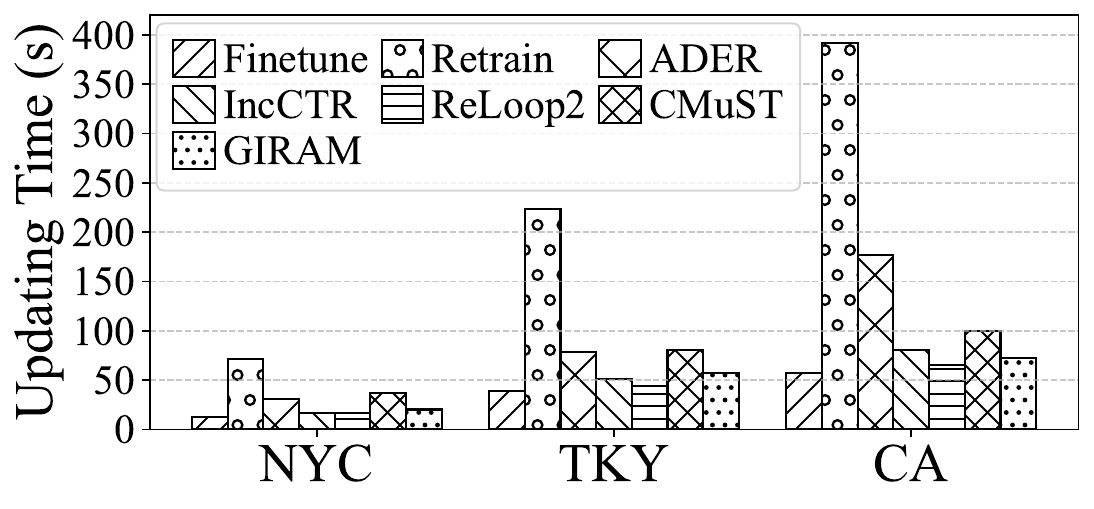}
\vspace{-10pt} 
\label{fig:time_GETNext}
\end{minipage}}
\subfigure[{DiffPOI.}]{
\begin{minipage}{0.28\linewidth}
\centering
\vspace{-2pt}
\includegraphics[width=1\linewidth]{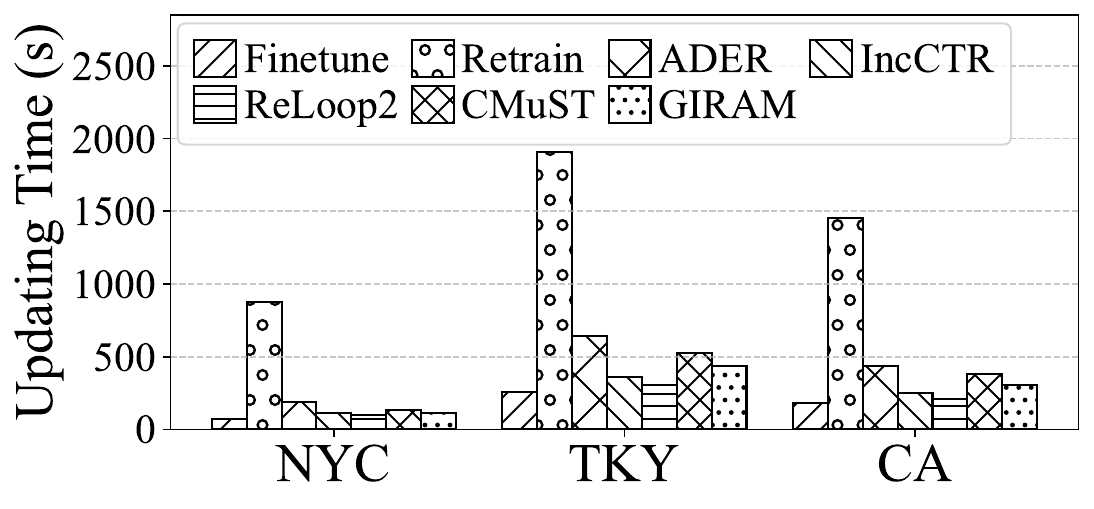}
\vspace{-10pt} 
\label{fig:time_DiffPOI}
\end{minipage}}
\vspace{-8pt}
\caption{{Updating time comparison using three backbones.}}
\label{fig:time}
\end{figure*}

Figure~\ref{fig:time} shows the average update time per block for three backbones, with the number of training epochs fixed at 10 to ensure a fair comparison. Finetuning is the fastest method, as it only trains on new data. Retraining is the slowest method, as it rebuilds the model from scratch. ADER is also costly due to hidden-state trajectory computation and replay sample selection, which requires repeated NPR calls. IncCTR and ReLoop2 are slightly slower than Finetune, but faster than the other methods since they reuse prior knowledge without complex selection. CMuST incurs extra overhead from contextual integration. GIRAM adds a small additional cost to Finetune for interest memory updates, but it consistently matches or surpasses the performance of all baselines while remaining far more efficient. Overall, GIRAM achieves continual updates with minimal overhead, effectively balancing accuracy and efficiency.

\subsubsection{Memory Usage}
\begin{figure*}[!t]
\centering
\subfigure[{Flashback.}] {
\begin{minipage}{0.28\linewidth}
\centering
\vspace{-10pt}
\includegraphics[width=1\linewidth]{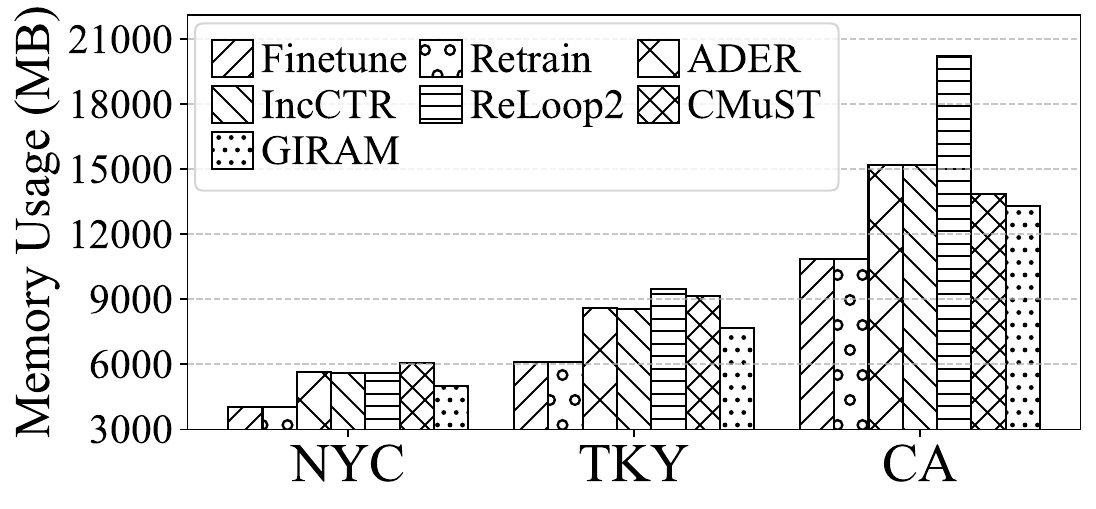}
\vspace{-10pt} 
\label{fig:memory_Flashback}
\end{minipage}}
\subfigure[{GETNext.}]{
\begin{minipage}{0.28\linewidth}
\centering
\vspace{-10pt}
\includegraphics[width=1\linewidth]{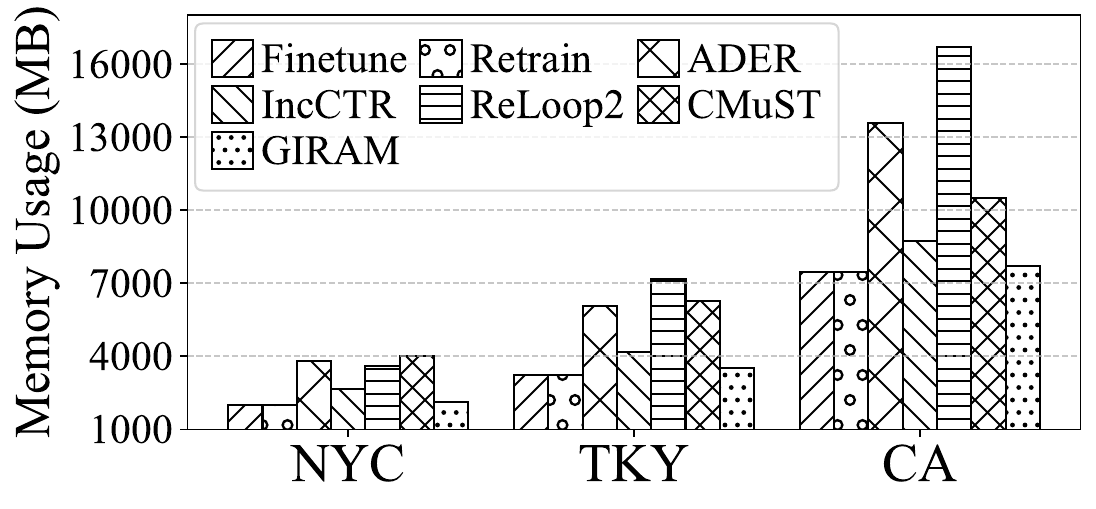}
\vspace{-10pt} 
\label{fig:memory_GETNext}
\end{minipage}}
\subfigure[{DiffPOI.}]{
\begin{minipage}{0.28\linewidth}
\centering
\vspace{-10pt}
\includegraphics[width=1\linewidth]{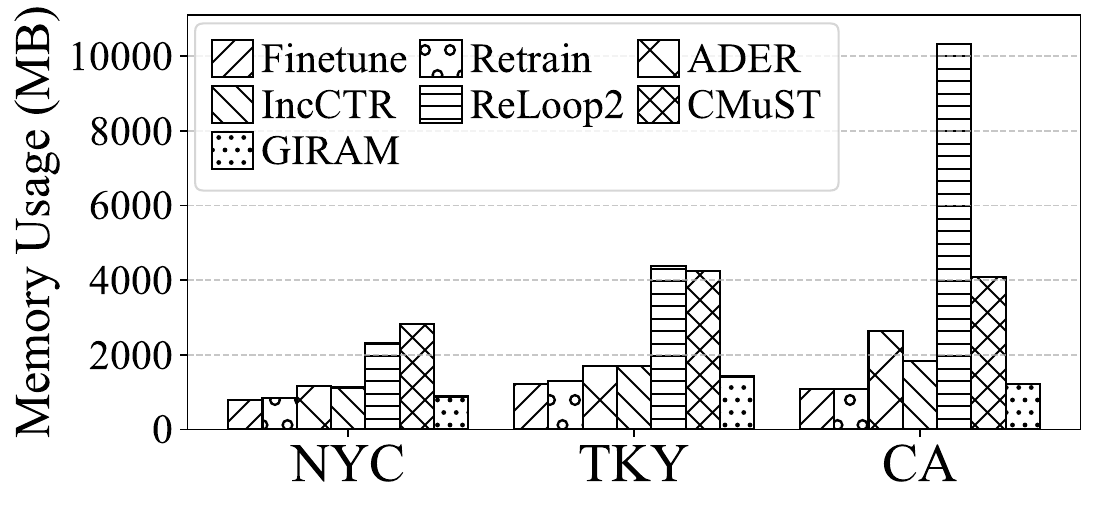}
\vspace{-10pt} 
\label{fig:memory_DiffPOI}
\end{minipage}}
\vspace{-8pt}
\caption{{Memory usage comparison using three backbones.}}
\vspace{-16pt}
\label{fig:memory_usage}
\end{figure*}

We further evaluate memory efficiency. As shown in Figure~\ref{fig:memory_usage}, Finetune and Retrain show the lowest usage, as memory is dominated by the backbone without auxiliary modules. Among continual methods, GIRAM is the most efficient thanks to its interest memory, which stores a fixed number of sustained interests with an adaptive update mechanism. ADER consumes more memory due to exemplar replay, which computes and stores hidden states for sample selection. IncCTR requires more memory than GIRAM because its knowledge distillation introduces an auxiliary loss for aligning teacher–student outputs. Despite using sparse POI vectors like GIRAM, ReLoop2 is the most memory-intensive because its error memory and per-user hash functions scale with the number of users, resulting in a high cost, particularly on the user-rich CA dataset. CMuST also uses extra memory due to its Multi-dimensional Spatio-Temporal Interaction module. Overall, GIRAM achieves the highest recommendation accuracy while remaining memory-efficient.

\vspace{-1pt}
\subsection{Ablation Study}
\vspace{-1pt}
To assess the contribution of each component in GIRAM, we design five variants: 
(1) \emph{w/o CKE} replaces the context-aware key encoder with the backbone latent vector; 
(2) \emph{w/o GKR} removes the generative key-based retrieval and instead retrieves the value of the most similar key; 
(3) \emph{w/o CS} substitutes the consistency-aware scoring with fixed base values for interest update and fusion; 
(4) \emph{w/o SI} discards sustained interests and relies only on recent ones; 
(5) \emph{w/o RI} discards recent interests and uses only sustained ones.

We report the average Acc@5 and MRR on three datasets (Table~\ref{tb:ablation}). GIRAM consistently outperforms all variants, confirming the effectiveness of its core modules. The drop in \emph{w/o CKE} highlights the importance of encoding contextual factors (location, time, POI category) for continual recommendation. \emph{w/o GKR} shows that generating multiple candidate keys improves retrieval accuracy by identifying more relevant sustained interests. The decline in \emph{w/o CS} demonstrates that adaptive consistency scoring is crucial for updating memory and balancing sustained and recent interests. Results of \emph{w/o SI} and \emph{w/o RI} further indicate that both sustained and recent interests are indispensable; removing either prevents effective integration of diverse interest patterns.

{We also compare the conditional VAE-based key generator with variants based on autoencoders and GANs. As shown in Table~\ref{tb:ablation}, GIRAM (with conditional VAE) achieves superior performance in all cases. Autoencoders struggle to produce diverse and independent keys due to their deterministic nature, resulting in outcomes similar to single-key retrieval. GANs suffer from instability and mode collapse, resulting in unreliable key representations. In contrast, the conditional VAE ensures stable training and diverse key generation, leading to superior performance.}

\begin{table}[!tb]\scriptsize
\setlength{\tabcolsep}{1.1mm}
\renewcommand{\arraystretch}{0.82}
\centering
\vspace{-4pt}
\caption{The results of ablation studies.}\label{tb:ablation}
\vspace{-4pt}
\begin{tabular}{c|l|cc|cc|cc}
\hline
\multirow{2}*{Backbones} & \multirow{2}*{Variants} & \multicolumn{2}{c|}{NYC} & \multicolumn{2}{c|}{TKY} & \multicolumn{2}{c}{CA}\\
\cline{3-8}
~& & Acc@5 & MRR & Acc@5 & MRR & Acc@5 & MRR\\
\hline
\multirow{8}*{Flashback} & (1) \emph{w/o CKE} & 0.4750 & 0.3465 & 0.4272 & 0.3082 & 0.2518 & 0.1801\\
~& (2) \emph{w/o GKR} & 0.4848 & 0.3456 & 0.4305 & 0.3071 & 0.2620 & 0.1866\\
~& (3) \emph{w/o CS} & 0.4869 & 0.3489 & 0.4466 & 0.3203 & 0.2670 & 0.1918\\
~& (4) \emph{w/o SI} & 0.4704 & 0.3354 & 0.4271 & 0.3054 & 0.2473 & 0.1828\\
~& (5) \emph{w/o RI} & 0.4757 & 0.3357 & 0.4395 & 0.3108 & 0.2535 & 0.1876\\
~& {GIRAM-AE}  & {0.4909} & {0.3466} & {0.4397} & {0.3072} & {0.2602} & {0.1922} \\
~& {GIRAM-GAN} & {0.4930} & {0.3505} & {0.4569} & {0.3243} & {0.2708} & {0.1958} \\
~& \textbf{GIRAM} & \textbf{0.5030} & \textbf{0.3610} & \textbf{0.4618} & \textbf{0.3331} & \textbf{0.2773} & \textbf{0.2028}\\
\hline
\multirow{8}*{GETNext} & (1) \emph{w/o CKE} & 0.4564 & 0.3239 & 0.4171 & 0.3015 & 0.2385 & 0.1786\\
~& (2) \emph{w/o GKR} & 0.4741 & 0.3463 & 0.4345 & 0.3158 & 0.2391 & 0.1824\\
~& (3) \emph{w/o CS} & 0.4755 & 0.3452 & 0.4297 & 0.3127 & 0.2486 & 0.1850\\
~& (4) \emph{w/o SI} & 0.4587 & 0.3364 & 0.4153 & 0.2981 & 0.2444 & 0.1764\\
~& (5) \emph{w/o RI} & 0.4599 & 0.3278 & 0.4293 & 0.2953 & 0.2468 & 0.1774\\
~& {GIRAM-AE}  & {0.4759} & {0.3513} & {0.4418} & {0.3182} & {0.2438} & {0.1908} \\
~& {GIRAM-GAN} & {0.4817} & {0.3504} & {0.4339} & {0.3121} & {0.2354} & {0.1847} \\
~& \textbf{GIRAM} & \textbf{0.4877} & \textbf{0.3535} & \textbf{0.4454} & \textbf{0.3234} & \textbf{0.2589} & \textbf{0.1920}\\
\hline
\multirow{8}*{DiffPOI} & (1) \emph{w/o CKE} & 0.4694 & 0.4258 & 0.4525 & 0.4153 & 0.2508 & 0.2155\\
~& (2) \emph{w/o GKR} & 0.4959 & 0.4591 & 0.4816 & 0.4340 & 0.2699 & 0.2337\\
~& (3) \emph{w/o CS} & 0.5008 & 0.4584 & 0.4873 & 0.4382 & 0.2782 & 0.2375\\
~& (4) \emph{w/o SI} & 0.3866 & 0.3554 & 0.4124 & 0.3657 & 0.2318 & 0.2005\\
~& (5) \emph{w/o RI} & 0.4902 & 0.4516 & 0.4773 & 0.4233 & 0.2636 & 0.2377\\
~& {GIRAM-AE}  & {0.5038} & {0.4628} & {0.4886} & {0.4343} & {0.2722} & {0.2352} \\
~& {GIRAM-GAN} & {0.5050} & {0.4677} & {0.4928} & {0.4471} & {0.2806} & {0.2403} \\
~& \textbf{GIRAM} & \textbf{0.5107} & \textbf{0.4697} & \textbf{0.5030} & \textbf{0.4536} & \textbf{0.2882} & \textbf{0.2492}\\
\hline
\end{tabular}
\vspace{-16pt}
\end{table}

\begin{figure}[!t]
\centering
\hspace{-20pt}
\subfigure[{NYC dataset.}]{
    \includegraphics[width=0.33\linewidth]{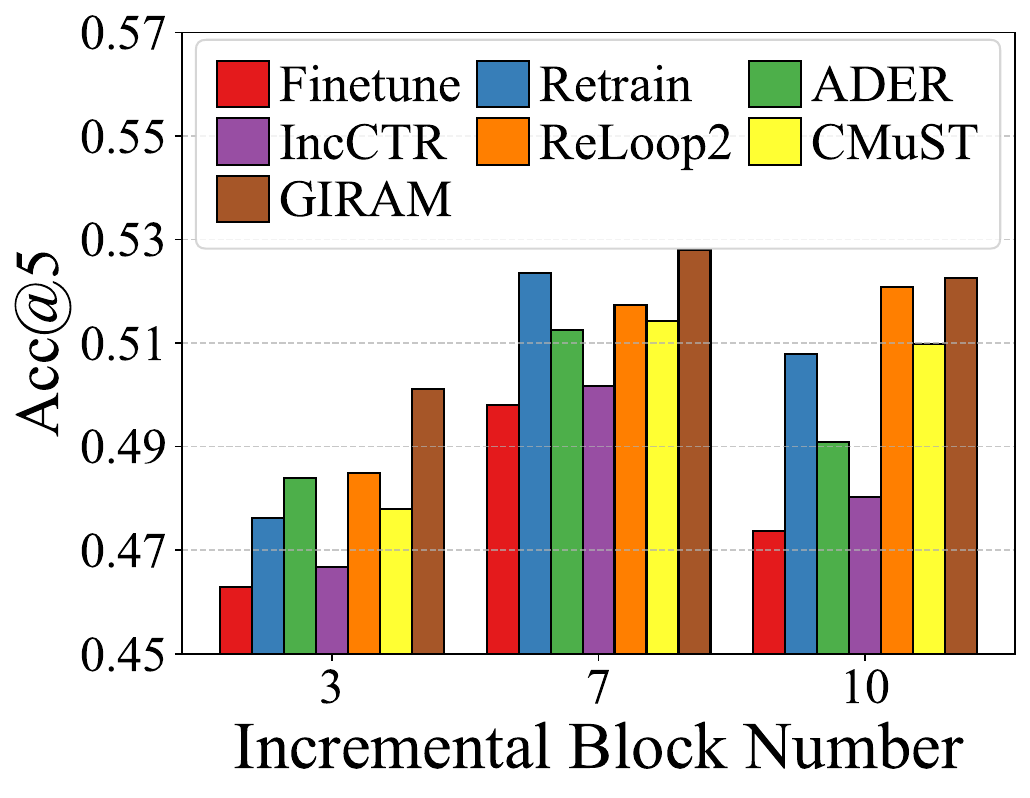}}
\hspace{-9pt}
\subfigure[{TKY dataset.}]{
    \includegraphics[width=0.33\linewidth]{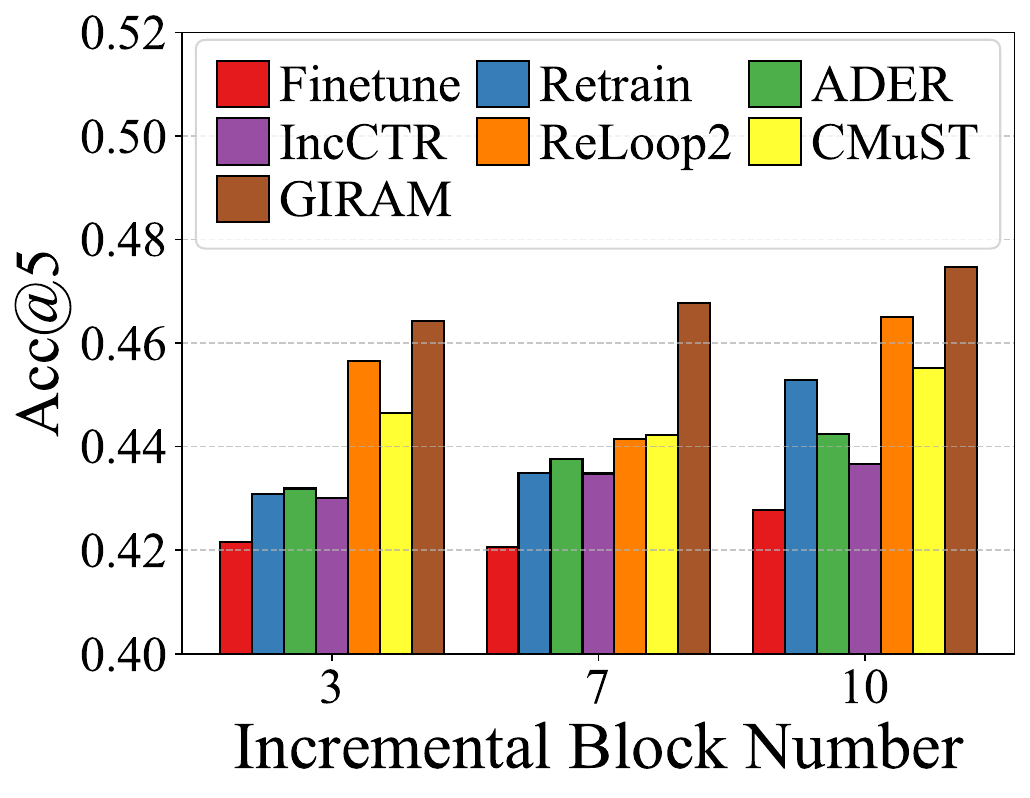}}
\hspace{-9pt}
\subfigure[{CA dataset.}]{
    \includegraphics[width=0.33\linewidth]{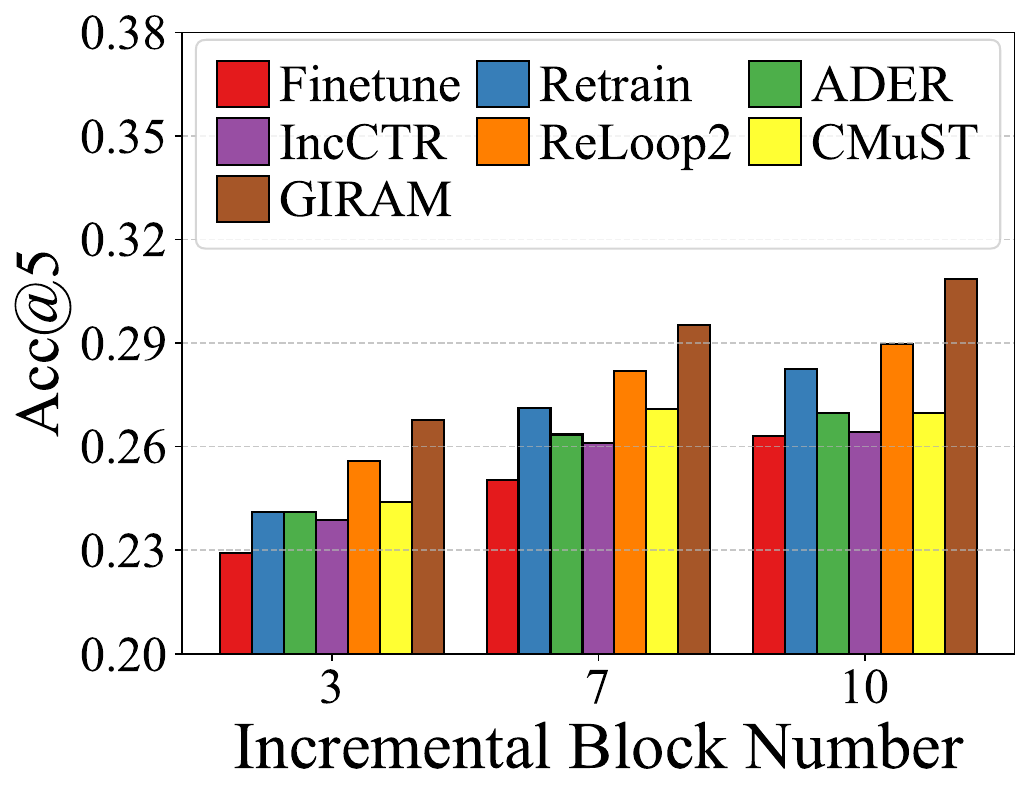}}
    \hspace{-9pt}
\vspace{-8pt}
\caption{{Results with different incremental block sizes (Flashback backbone).}}
\vspace{-15pt}
\label{fig:size}
\end{figure}

\vspace{-6pt}
{
\subsection{Performance Across Different Block Sizes}
To further assess the robustness of GIRAM under different incremental block settings, we conduct experiments using Flashback as the backbone, where the block sizes are set to 3, 7, and 10 (in addition to the original setting of 5). We report the average Acc@5 on each test block. As Figure~\ref{fig:size} shows, GIRAM outperforms the baselines consistently across all block sizes, demonstrating its effectiveness in capturing evolving user preferences.}

\subsection{Hyperparameter Sensitivity Analysis}
\begin{figure}[!t]
\centering
\vspace{-8pt}
\includegraphics[width=0.95\linewidth]{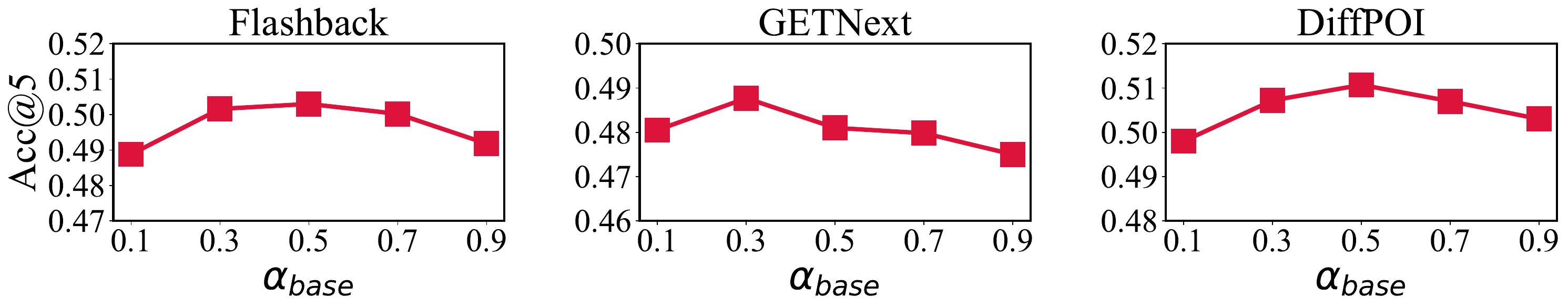}
\vspace{-10pt}
\caption{Performance comparison of different $\alpha_{\text{base}}$.}\label{fig:alpha}
\vspace{-8pt}
\end{figure}

\begin{figure}[!t]
\centering
\includegraphics[width=0.95\linewidth]{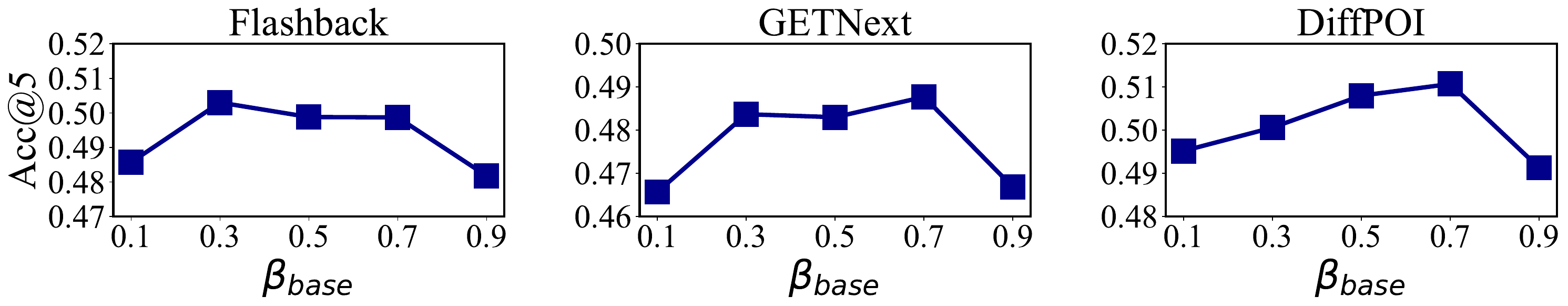}
\vspace{-10pt}
\caption{Performance comparison of different $\beta_{\text{base}}$.}\label{fig:beta}
\vspace{-8pt}
\end{figure}

\begin{figure}[!t]
\centering
\includegraphics[width=0.95\linewidth]{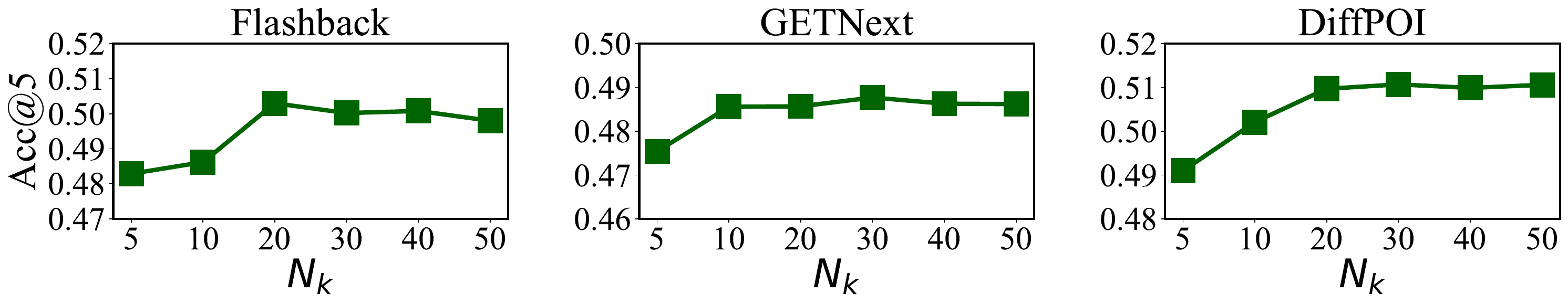}
\vspace{-10pt}
\caption{Performance comparison of different $N_k$.}\label{fig:key_num}
\vspace{-8pt}
\end{figure}

\begin{figure}[!t]
\centering
\includegraphics[width=0.95\linewidth]{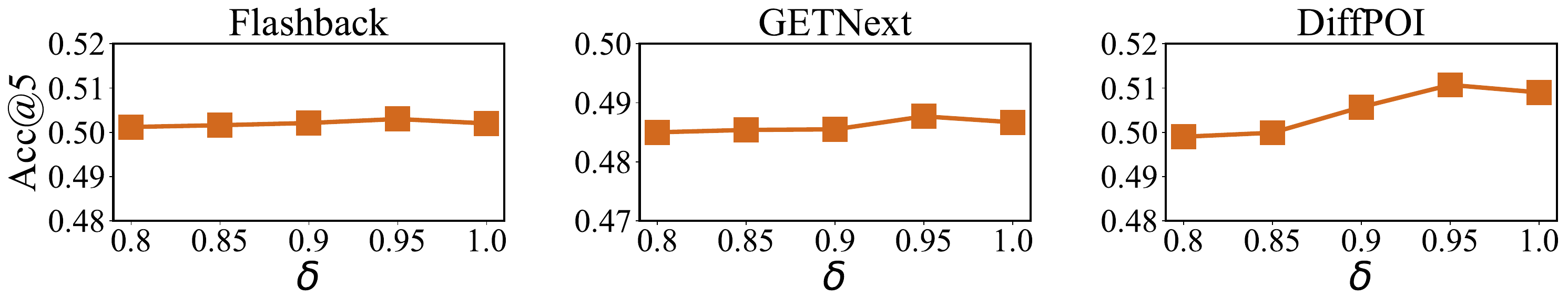}
\vspace{-10pt}
\caption{Performance comparison of different $\delta$.}\label{fig:delta}
\vspace{-8pt}
\end{figure}

\vspace{-1pt}
\label{sec:hyper}
A hyperparameter sensitivity analysis is conducted on NYC dataset to examine the effect of key parameters on GIRAM. Specifically, we report the average Acc@5 on three backbones by varying four parameters: update weight $\alpha_{\text{base}}$, fusion weight $\beta_{\text{base}}$, number of generated keys $N_k$, and update threshold $\delta$. We vary $\alpha_{\text{base}}$ and $\beta_{\text{base}}$ from 0.1 to 0.9, $N_k$ from 5 to 50, and $\delta$ from 0.8 to 1.0. The results are shown in Figures~\ref{fig:alpha}–\ref{fig:delta}.

For $\alpha_{\text{base}}$, GIRAM achieves the best Acc@5 at 0.5 with Flashback and DiffPOI, and at 0.3 with GETNext. For $\beta_{\text{base}}$, the best values are 0.3 for Flashback and 0.7 for GETNext and DiffPOI. Notably, $\alpha_{\text{base}} = \beta_{\text{base}} = 0.5$ provides near-optimal performance across all backbones, suggesting a robust default. These results indicate that extreme $\alpha_{\text{base}}$ or $\beta_{\text{base}}$ disrupts the balance between sustained and recent interests, harming recommendation quality. For $N_k$, the optimal is 20 with Flashback and 30 with GETNext and DiffPOI: too few keys limit preference diversity, while too many introduce noise. The best $\delta$ is 0.95, consistent with the empirical similarity distribution between query keys and memory entries, ensuring reliable updates while avoiding false matches.

\begin{figure}[!t]
\centering
\includegraphics[width=0.95\linewidth]{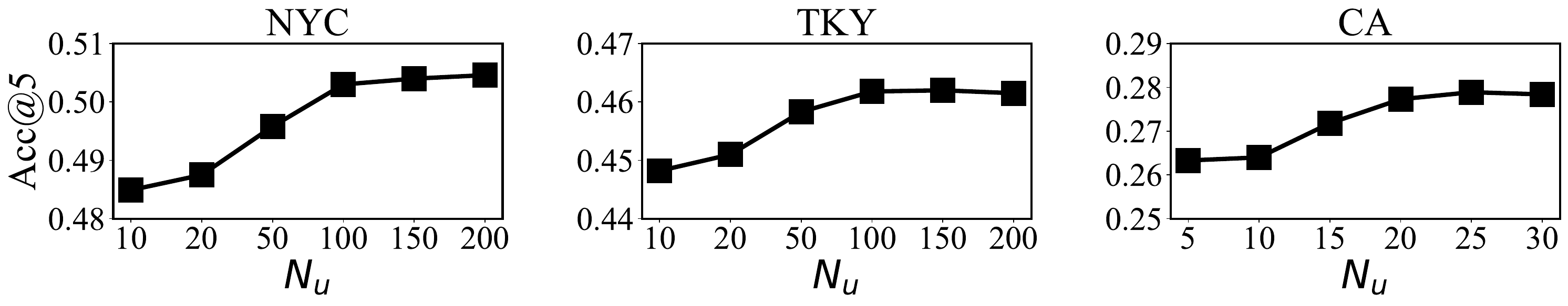}
\vspace{-10pt}
\caption{{Performance comparison of different memory sizes $N_u$.}}\label{fig:memory_size}
\vspace{-12pt}
\end{figure}

{We further analyze the interest memory size across datasets, using Flashback as the backbone. The results in Figure~\ref{fig:memory_size} show that the average Acc@5 saturates at $N_u=50$ for both NYC and TKY and at $N_u=20$ for CA, confirming the need for dataset-specific interest memory sizes.}

\vspace{-2pt}
\subsection{Case Study}
\begin{figure}[!t]
\centering
\vspace{-8pt}
\includegraphics[width=0.85\linewidth]{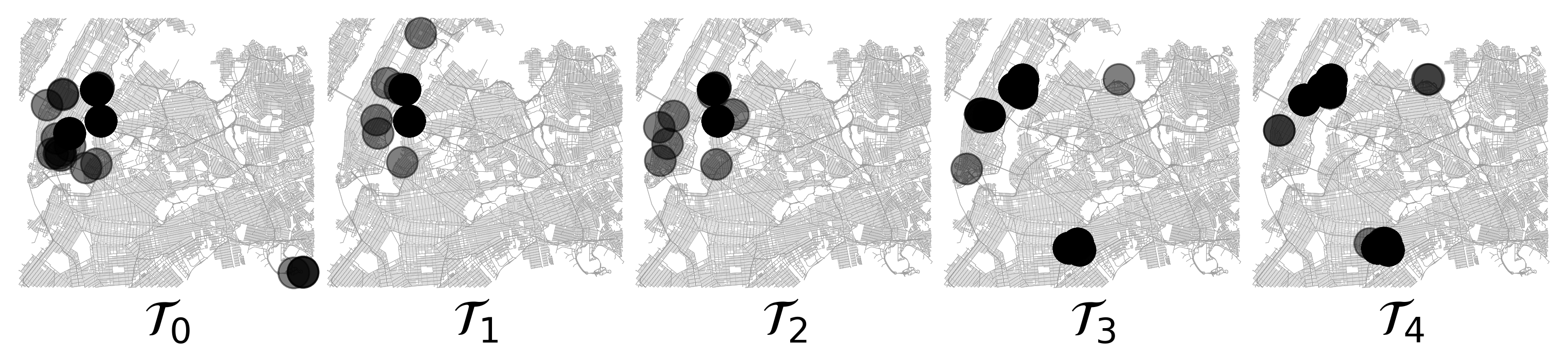}
\vspace{-11pt}
\caption{Visualization of historical POIs of User \#137.}\label{fig:location}
\vspace{-11pt}
\end{figure}

\begin{figure}[!t]
\centering
\includegraphics[width=0.85\linewidth]{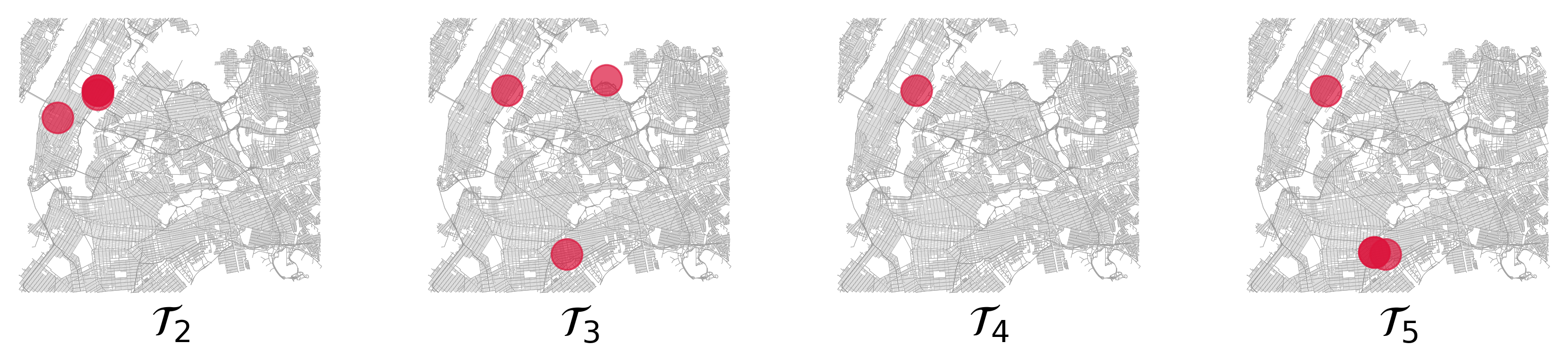}
\vspace{-11pt}
\caption{Visualization of ground truth POIs of User \#137.}\label{fig:real}
\vspace{-11pt}
\end{figure}

\begin{figure}[!t]
\centering
\includegraphics[width=0.85\linewidth]{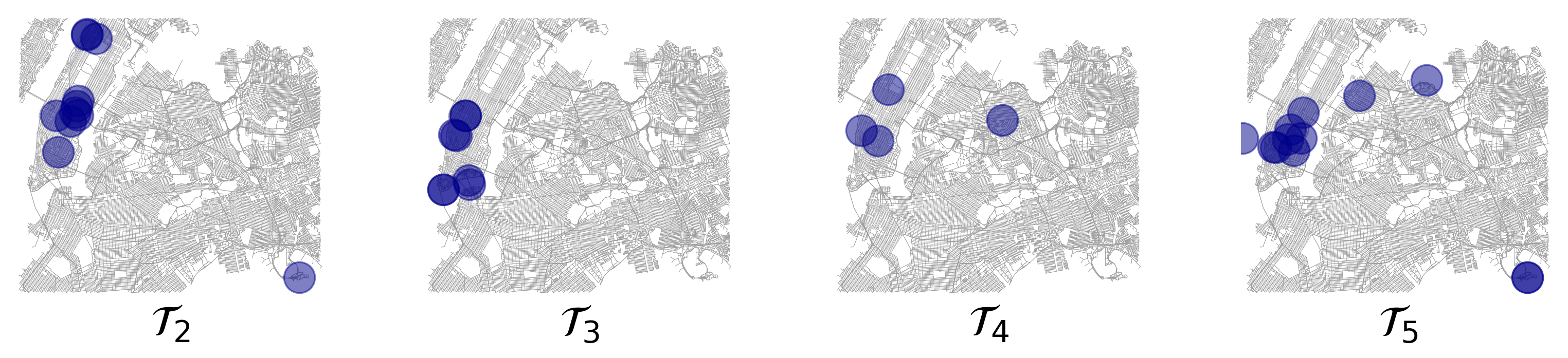}
\vspace{-11pt}
\caption{Visualization of top-5 recommendations of User \#137 using Static.}\label{fig:static}
\vspace{-11pt}
\end{figure}

\begin{figure}[!t]
\centering
\includegraphics[width=0.85\linewidth]{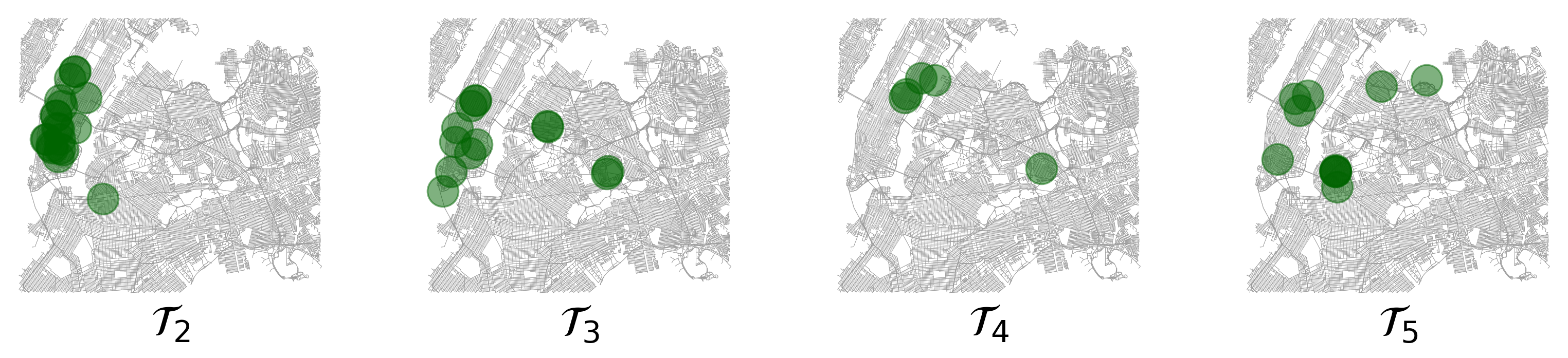}
\vspace{-11pt}
\caption{Visualization of top-5 recommendations of User \#137 using Finetune.}\label{fig:finetune}
\vspace{-11pt}
\end{figure}

\begin{figure}[!t]
\centering
\includegraphics[width=0.85\linewidth]{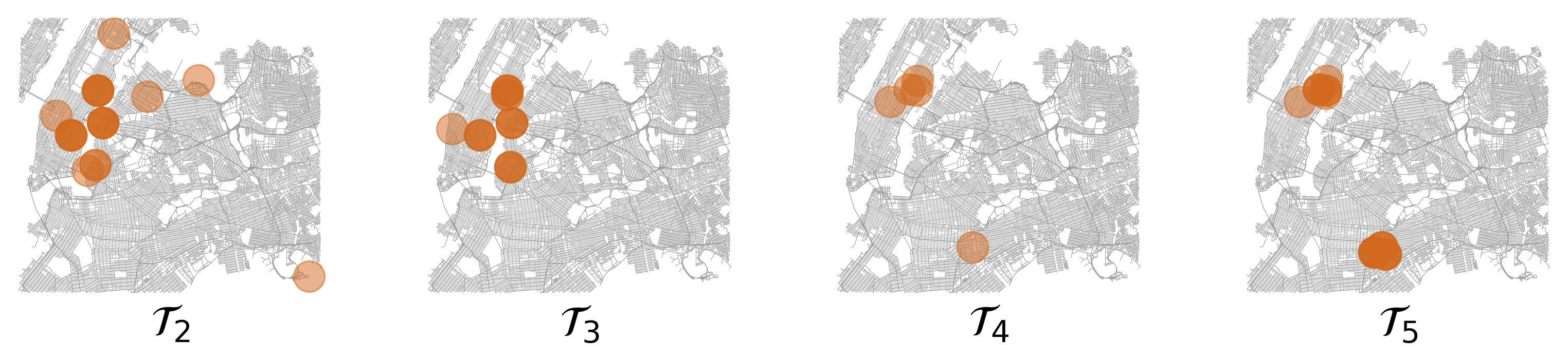}
\vspace{-11pt}
\caption{Visualization of top-5 recommendations of User \#137 using GIRAM.}\label{fig:memory}
\vspace{-14pt}
\end{figure}

\vspace{-2pt}
To further illustrate how GIRAM captures evolving user interests over time, we conduct a case study on User~\#137. Figure~\ref{fig:location} visualizes the user’s historical POIs from block $\mathcal{T}_0$ to block $\mathcal{T}_4$, while Figure~\ref{fig:real} shows the ground truth POIs from block $\mathcal{T}_2$ to block $\mathcal{T}_5$. Figure~\ref{fig:static}, Figure~\ref{fig:finetune}, and Figure~\ref{fig:memory} display the top-5 recommendations produced by Static, Finetune, and GIRAM, respectively. The user’s check-in behavior evolves across temporal blocks, indicating notable shifts in interest and location. Overall, we can find that GIRAM generates significantly more accurate recommendations than both Static and Finetune.

To provide a more detailed analysis, we focus on block $\mathcal{T}_5$ as a representative example. The ground truth POIs include: \#1461: (40.7670, -73.9579), Gym / Fitness Center; \#2746: (40.6615, -73.9163), Subway Station; \#5099: (40.6602, -73.9065), Home. The Static model continues to recommend POIs that are frequently visited during block $\mathcal{T}_0$, ignoring more recent behavioral changes. As shown in Figure~\ref{fig:static}, the spatial distribution of its predictions aligns closely with the historical POIs in block $\mathcal{T}_0$ in Figure~\ref{fig:location}, which reflects the model’s inability to adapt to evolving user preferences. The Finetune model updates based on the most recent data and recommends POIs that appear frequently in blocks $\mathcal{T}_3$ and $\mathcal{T}_4$. However, due to its disregard for long-term historical information, it fails to identify the user’s sustained interests. As a result, none of its predictions in block $\mathcal{T}_5$ match the actual visited POIs in block $\mathcal{T}_5$ (Figure~\ref{fig:finetune}). In contrast, GIRAM accurately recommends all three ground truth POIs in block $\mathcal{T}_5$ (Figure~\ref{fig:memory}). This demonstrates its strength in modeling both short-term contextual relevance and long-term interest continuity. By retrieving relevant keys from its memory conditioned on the current trajectory context, GIRAM can synthesize sustained and recent interests to make accurate recommendations.
\section{CONCLUSION}
In this work, we explore the novel problem of continual next POI recommendation, which focuses on dynamically updating recommendation models to adapt to evolving user preferences without full retraining. To address this challenge, we propose GIRAM, a flexible and model-agnostic framework that integrates four key components: interest memory, context-aware key encoding, generative key-based retrieval, and adaptive interest update and fusion. These components collaboratively enable efficient interest memory updates and maintain a balanced representation of both sustained and recent user preferences. Notably, GIRAM can be seamlessly integrated into a wide range of existing NPR models without architectural modifications. Extensive experiments on three real-world datasets demonstrate that GIRAM consistently outperforms state-of-the-art baselines in both recommendation accuracy and adaptability, while achieving high efficiency in both updating time and memory usage. Furthermore, ablation studies confirm the effectiveness of each component, highlighting GIRAM’s capability to model dynamic and evolving user interests in continual recommendation settings.

\section{ACKNOWLEDGMENT}
This paper was supported by the National Key R\&D Program of China 2024YFE0111800, NSFC U25B2049, NSFC U22B2037, NSFC U21B2046, and NSFC 62032001.

\section{AI-Generated Content Acknowledgement}
ChatGPT was used to assist with language polishing and improving readability in Section~\ref{sec:related_work}. All scientific content was created and verified by the authors. No figures, images, or code in this paper were generated by AI tools.

\bibliographystyle{IEEEtran}
\bibliography{references}

\end{document}